\documentclass{article}
\usepackage[margin=1.0in]{geometry}
\usepackage{amsmath}
\usepackage{amsthm}
\usepackage{amsfonts}
\usepackage{amssymb}
\usepackage{float}
\usepackage{graphicx}
\usepackage{caption}
\usepackage{algorithm}
\usepackage{algpseudocode}
\usepackage{subcaption}
\usepackage{url}
\usepackage{titling}
\usepackage{natbib}
\usepackage{appendix}
\usepackage{hyperref}
\usepackage{setspace}

\hypersetup{
     colorlinks=true,
     linkcolor=black,
     filecolor=blue,
     citecolor = black,      
     urlcolor=black,
}

\providecommand{\keywords}[1]{\textbf{\textit{Keywords:}} #1}

\def\y{\mathbf{y}}
\def\X{\mathbf{X}}

\makeatletter
\newcommand*\bigcdot{\mathpalette\bigcdot@{.5}}
\newcommand*\bigcdot@[2]{\mathbin{\vcenter{\hbox{\scalebox{#2}{$\m@th#1\bullet$}}}}}
\makeatother

\title{Bayesian Simultaneous Factorization and Prediction Using Multi-Omic Data}

\author{Sarah Samorodnitsky$^{a,\ast}$, Chris H. Wendt$^{b}$, Eric F. Lock$^{a}$\\
\small $^{a}$Division of Biostatistics, University of Minnesota, 420 Delaware St SE, Minneapolis, MN, USA 55455 \\
\small $^b$Minneapolis VA Health Care System, One Veterans Drive, Minneapolis, MN, USA 55417 \\
\small $^{*}$Corresponding author: Sarah Samorodnitsky; \tt{samor007@umn.edu}}
\date{}

\begin{document}



\maketitle

\begin{abstract}
\noindent
Understanding of the pathophysiology of obstructive lung disease (OLD) is limited by available methods to examine the relationship between multi-omic molecular phenomena and clinical outcomes. Integrative factorization methods for multi-omic data can reveal latent patterns of variation describing important biological signal. However, most methods do not provide a framework for inference on the estimated factorization, simultaneously predict important disease phenotypes or clinical outcomes, nor accommodate multiple imputation. To address these gaps, we propose Bayesian Simultaneous Factorization (BSF). We use conjugate normal priors and show that the posterior mode of this model can be estimated by solving a structured nuclear norm-penalized objective that also achieves rank selection and motivates the choice of hyperparameters. We then extend BSF to simultaneously predict a continuous or binary response, termed Bayesian Simultaneous Factorization and Prediction (BSFP). BSF and BSFP accommodate concurrent imputation and full posterior inference for missing data, including “blockwise” missingness, and BSFP offers prediction of unobserved outcomes. We show via simulation that BSFP is competitive in recovering latent variation structure, as well as the importance of propagating uncertainty from the estimated factorization to prediction. We also study the imputation performance of BSF via simulation under missing-at-random and missing-not-at-random assumptions. Lastly, we use BSFP to predict lung function based on the bronchoalveolar lavage metabolome and proteome from a study of HIV-associated OLD. Our analysis reveals a distinct cluster of patients with OLD driven by shared metabolomic and proteomic expression patterns, as well as multi-omic patterns related to lung function decline. Software is freely available at \url{https://github.com/sarahsamorodnitsky/BSFP}.
\end{abstract}

\doublespacing

\section{Introduction}\label{s:intro}


Despite increasing usage of combination antiretroviral therapies, obstructive lung disease (OLD) remains a frequent comorbidity among individuals living with HIV \citep{hirani2011prevalence}. We do not have a complete understanding of the factors associated with risk of developing OLD in this population. 
This motivates our use of multiple sources of omics data (multi-omic data), collected from bronchoalveolar lavage (lung) fluid in patients with HIV, to characterize the OLD pathophysiology. 
Our interests are two-fold: (1) to characterize patterns of molecular variation in lung fluid within this unique cohort, and (2) to relate molecular expression profiles to clinical measurements of lung function. 
Despite a growing body of statistical methods for multi-omics analysis, our HIV-OLD application and similar studies motivate new methodology that can simultaneously (1) identify latent components that explain variation within or across multi-omics data, (2) use the latent components to predict an outcome, (3) perform missing data imputation, and (4) fully characterize uncertainty in the factorization components, imputed data, and outcome predictions.  

There are several examples of exploratory integrative factorization methods, both non-Bayesian \citep{shen2012integrative, lock2013joint, zhou2015group, yang2016non, feng2018angle, gaynanova2019structural}, and Bayesian \citep{klami2013bayesian, chekouo2017bayesian, argelaguet2018multi}, which can be used to identify associations between sources in the form of low-rank structured variation.
A less-explored area is simultaneous factorization and prediction, which may reveal latent variation structure that is also related to an outcome or phenotype. 
We can incorporate prediction into existing exploratory methods in two steps: (1) identify a small number of latent components explaining variation in the data, and (2) use these as covariates in a prediction model. This was described in \cite{kaplan2017prediction} for multi-omic data, in \cite{samorodnitsky2022hierarchical} for multi-omic and multi-cohort data, and in \cite{hellton2016integrative} for clustering. Recently, one-step simultaneous factorization and prediction procedures have been proposed \citep{zhang2021joint, li2021integrative, palzer2022sjive, safo2022sparse}, including some Bayesian approaches \citep{chekouo2020bayesian, white2021bayesian}. While these methods contribute to the growing body of literature on supervised factorization approaches, they do not provide a framework for inference on the underlying factorization and do not accommodate missing values in either the data sources or the outcome.

To fill these gaps, we propose Bayesian Simultaneous Factorization (BSF), which estimates a partitioned factorization consisting of joint structure (variation shared across sources) and individual structure (variation specific to each source). 
Our approach can be viewed as an extension of probabilistic matrix factorization (PMF) \citep{mnih2007probabilistic, salakhutdinov2008bayesian} with a Gaussian likelihood and and conjugate Gaussian priors on the factorization components.  The posterior mode is the solution to a structured nuclear-norm penalized objective, which matches \cite{park2020integrative}'s UNIFAC decomposition. Solving this objective achieves rank selection, motivates our choice of prior hyperparameters, and allows efficient initialization of a Gibbs sampling algorithm to estimate the posterior distributions of the factorization parameters. We also propose Bayesian Simultaneous Factorization and Prediction (BSFP), which extends BSF by using factors driving joint and individual structures to predict an outcome. We focus on a continuous outcome, but this is naturally extended to accommodate any Bayesian predictive model and we describe our implementation for a binary outcome in the Appendix. Both BSF and BSFP offer full posterior inference and can be used for multiple imputation of missing data from the posterior predictive distribution, including in ``blockwise" missing scenarios in which an entire sample from a source is unavailable.

The remainder of our article is organized as follows: in Section~\ref{s:Methods}, we review PMF, UNIFAC, and introduce BSF and BSFP. In Section~\ref{s:Simulation}, we compare BSFP to existing one- and two-step approaches to factorization and prediction via simulation and assess the imputation accuracy of BSFP against other approaches. In Section~\ref{s:data_application}, we describe applying BSFP to predict lung function using proteomic and metabolomic data to study HIV-associated OLD. We conclude with a discussion of the method and potential new directions. 

\section{Methods}\label{s:Methods}

\subsection{Notation}\label{ss:notation}
We first introduce notation used throughout. Bold, uppercase letters, e.g. $\X$, denote matrices. Bold, lowercase letters, e.g. $\y$, denote vectors. Unbolded uppercase and lowercase letters, e.g. $R$ and $q$, denote scalars. For illustration, 
consider $q$ data sources, e.g. omics datasets, measured on $n$ samples. Let $\X_s:p_s \times n$ represent source $s$, containing $p_s$ biomarkers, oriented such that the columns are the samples. $\X_{1}, \dots, \X_{q}$ are linked by their columns, meaning they contain biomarkers measured on a shared set of $n$ samples. We use $\X_s[j,i]$, $j=1,\dots, p_s$ and $i=1,\dots, n$ to denote expression of the $j$th feature by the $i$th sample in source $s$. Similarly, $\X_{s}[j,\cdot]$ represents the $j$th feature across all samples (the $j$th row of $\X_{s}$) and $\X_{s}[\cdot,i]$ represents the expression values for the $i$th sample (the $i$th column of $\X_{s}$). We use subscripts and unbolded letters to index elements within a vector, e.g., 
$y_i$ is the $i$th entry in $\y$. We let $\X_{\bigcdot} = \begin{pmatrix} \X_{1}^{T} & \dots & \X_{q}^{T} \end{pmatrix}^{T}$ denote the column-concatenated, full data matrix containing biomarkers from all $q$ sources such that $R = \hbox{rank}(\X_{\bigcdot})$. Let $p=\sum_{s=1}^{q} p_s$ represent the total number of observed biomarkers. We define the squared Frobenius norm, $||\cdot||^2_F$, of $\X_s$ as $\sum_{j=1}^{p_s} \sum_{i=1}^n \X_s[j,i]^2$, i.e. the sum of squared entries in $\X_s$, and the nuclear norm, $||\cdot||_*$, of $\X_s$ as $\sum_{k=1}^{R_s} \sigma_k(\X_s)$ where $\sigma_k(\X_s)$ denotes the $k$th singular value of $\X_s$ and $R_s = \hbox{rank}(\X_s)$. 

\subsection{Review of Probabilistic Matrix Factorization (PMF)}\label{ss:PMF}

Before describing our proposed method in Section~\ref{ss:Proposed_Model}, we review probabilistic matrix factorization (PMF) in this section and UNIFAC in Section~\ref{ss:UNIFAC}. Consider a single real-valued matrix, $\X\in \mathbb{R}^{p \times n}$. 
PMF is a 
Bayesian linear factor model, in which observations in $\X$ are assumed to be driven by a small number, $r < \hbox{rank}(\X)$, of latent factors or components. These latent factors are contained in a matrix $\mathbf{V}\in\mathbb{R}^{n\times r}$, termed \emph{scores}, which are mapped to the space spanned by the observed features or biomarkers by matrix $\mathbf{U}\in\mathbb{R}^{p\times r}$, termed \emph{loadings}. Assuming $\X = \mathbf{U}\mathbf{V}^T + \mathbf{E}$, then $\mathbf{U}\mathbf{V}^T : p\times n$ is a low-rank approximation to the observed $\X$ where $\hbox{Var}(\mathbf{E}[j,i]) = \sigma^2$ for $j=1,\dots, p$, $i=1,\dots, n$. We refer to $\mathbf{U}\mathbf{V}^T$ as a \emph{structure}, as it contains structured variation underlying $\X$. PMF imposes the following conditional likelihood on the observed entries in $\X$ given $\mathbf{U}$, $\mathbf{V}$, and $\sigma^2$:
\begin{equation}
\label{eq:PMF_Likelihood}
    \X|\mathbf{U}, \mathbf{V}, \sigma^2 \sim \prod_{j=1}^p \prod_{i=1}^n \hbox{Normal}\left(\X[j,i]\mid \mathbf{U}[j,\cdot] \mathbf{V}[i,\cdot]^T, \sigma^2 \right)
\end{equation}
where $\hbox{Normal}(\cdot|\cdot,\cdot)$ represents the density of the Gaussian distribution. \cite{mnih2007probabilistic} impose mean-zero, Gaussian priors on the factorization components, $\mathbf{U}$ and $\mathbf{V}$:
\begin{align}
\begin{split}
    \mathbf{U}|\sigma^2_U &\sim \prod_{j=1}^p \hbox{Normal}\left( \mathbf{U}[j,] \mid 0, \sigma^2_U \mathbf{I}_{r\times r} \right) \\
    \mathbf{V}|\sigma^2_V &\sim \prod_{i=1}^n \hbox{Normal}\left( \mathbf{V}[i,] \mid 0, \sigma^2_V \mathbf{I}_{r\times r} \right)
\end{split}
\end{align}
where $\mathbf{I}_{r\times r}$ is an $r \times r$ identity matrix and $\sigma^2_U, \sigma^2_V > 0$. 

\subsection{Review of UNIFAC}\label{ss:UNIFAC}

Whereas PMF applies to a single matrix, the UNIFAC method was developed as a simultaneous low-rank decomposition of multiple matrices (e.g., multi-omic data) using random matrix theory. Consider $q$ omics sources, $\X_1, \dots, \X_q$ as defined in Section~\ref{ss:notation}, row-centered to have mean 0. 
The UNIFAC decomposition is as follows:
\begin{align}
\begin{split}
    \X_{\bigcdot} &= \mathbf{S}_{\bigcdot} + \mathbf{E}_{\bigcdot} \\
    &= \mathbf{J}_{\bigcdot} + \mathbf{A}_{\bigcdot} + \mathbf{E}_{\bigcdot} \label{eq:joint.indiv.decomp}\\
    &= \mathbf{U}_{\bigcdot} \mathbf{V}^T + \mathbf{W}_{\bigcdot} \mathbf{V}_{\bigcdot} ^T + \mathbf{E}_{\bigcdot} \\
    &= \begin{pmatrix}
    \mathbf{U_1} \\ \vdots \\ \mathbf{U_q}
    \end{pmatrix} \mathbf{V}^T + \begin{pmatrix}
    \mathbf{W_1} & & \\
    & \ddots &\\
    & & \mathbf{W_q}
    \end{pmatrix} \begin{pmatrix}
    \mathbf{V_1}^T \\ \vdots \\ \mathbf{V_q}^T
    \end{pmatrix} + \begin{pmatrix}
    \mathbf{E_1} \\ \vdots \\ \mathbf{E_q}
    \end{pmatrix}
\end{split}
\end{align}
where $\mathbf{J}_1, \dots, \mathbf{J}_q$ are matrices of rank $r < R$ and $\mathbf{A}_1, \dots, \mathbf{A}_q$ are matrices of rank $r_s < R_s$. $\mathbf{J}_s$ contains joint structure, latent expression patterns shared by all sources, as it is reflected in source $s$. Individual structure, $\mathbf{A}_s$, contains latent expression patterns unique to source $s$. $\mathbf{E}_s$ reflects Gaussian noise with variance $\sigma^2_s$ not captured in the decomposition. $\mathbf{J}_s$ is decomposed into $\mathbf{V}:n \times r$, the joint scores, which contain latent factors expressed by the samples in all sources, and the joint loadings, $\mathbf{U}_s:p_s \times r$ which maps these factors to the observed biomarkers in source $s$. Similarly, $\mathbf{A}_s$ is decomposed into $\mathbf{V}_s: n\times r_s$, the individual scores, which contain the latent factors unique to each source, and the individual loadings, $\mathbf{W}_s: p_s \times r_s$, which map the factors to the feature space spanned by the biomarkers in source $s$. 
\cite{park2020integrative} (and the extension described in \cite{lock2022bidimensional}) propose estimating the UNIFAC decomposition in Equation~\ref{eq:joint.indiv.decomp}  by minimizing the 
following structured nuclear-norm penalized objective:
\begin{align}
\begin{split}
\label{eq:nn_obj}
\{\hat{\mathbf{J}}_s, \hat{\mathbf{A}}_s |s=1,\dots q\} &= \min_{\{\mathbf{J}_s, \mathbf{A}_s\}_{s=1}^q} \frac{1}{2} \sum_{s=1}^q || \X_s - \mathbf{J}_s - \mathbf{A}_s ||_F^2 + \lambda ||\mathbf{J}_{\bigcdot}||_* + \sum_{s=1}^q \lambda_s ||\mathbf{A}_s||_*
\end{split}
\end{align}
where $\hat{\mathbf{J}}_s = \hat{\mathbf{U}}_s \hat{\mathbf{V}}^T$ and $\hat{\mathbf{A}}_s = \hat{\mathbf{W}}_s \hat{\mathbf{V}}_s^T$. Equation~\ref{eq:nn_obj} is convex and minimized using an iterative soft singular value thresholding algorithm on the singular values of the structures. This procedure retains left- and right-singular vectors of $\mathbf{J}_s$ and $\mathbf{A}_s$ for which the corresponding singular values, $\sigma_k(\mathbf{J}_s) > 
\lambda, k=1,\dots, R$ and $\sigma_{k_s}(\mathbf{A}_s) > \lambda_s, k_s = 1,\dots, R_s$. Thus, rank selection for the joint and individual structures is a function of the tuning parameters, $\lambda$ and $\lambda_s$. Fixing $\lambda_s = \sqrt{n} + \sqrt{p_s}$ is a reasonable choice because it provides a tight upper bound on the largest singular value of the error, $\mathbf{E}_{s}$, assuming the sources have unit error variance,  $\sigma^2_s = 1$ \citep{rudelson2010non}. This effectively retains structure driven by components not attributed to error. This choice also meets the requirements established in \cite{park2020integrative} for a uniquely identifiable and non-zero decomposition, which we discuss further in Section~\ref{ss:identifiability}. The penalty on the joint structure, $\lambda$, is fixed to $\lambda = \sqrt{p} + \sqrt{n}$ by an analogous argument, as $\mathbf{E}_{\bigcdot}$ is also a mean-zero Gaussian random matrix. Prior to estimating the decomposition, we scale the sources to have unit error variance by dividing each by its estimated error standard deviation given by the median absolute deviation (MAD) estimator \citep{gavish2017optimal}. 


\subsection{Bayesian Simultaneous Factorization (BSF)}\label{ss:Proposed_Model}
The resulting factorization from UNIFAC is also the mode of a Bayesian posterior that naturally extends the PMF model, which we leverage for our Bayesian Simultaneous Factorization (BSF) model.  Theorem 1 in \cite{park2020integrative} establishes the equivalence between minimizing the nuclear norm objective \eqref{eq:nn_obj} and minimizing a similar objective with matrix-defined $L_2$ penalties (Frobenius norms) on the scores and loadings:
\begin{align}
\begin{split}
\label{eq:l2_penalized}
\{\hat{\mathbf{U}}_s, \hat{\mathbf{V}}, \hat{\mathbf{W}}_s, \hat{\mathbf{V}}_s | s=1,\dots q \} = \min_{\{\mathbf{U}_{s}, \mathbf{V}, \mathbf{W}_s, \mathbf{V}_s\}_{s=1}^q} \sum_{s=1}^q ||\X_s - \mathbf{U}_s \mathbf{V}^T - \mathbf{W}_s \mathbf{V_s}^T||_F^2 \\ + \lambda (||\mathbf{U}_{\bigcdot}||_F^2 + ||\mathbf{V} ||_F^2) + \sum_{s=1}^q \lambda_s (||\mathbf{W}_s ||_F^2 + || \mathbf{V}_s||_F^2).
\end{split}
\end{align}
Further, Equation~\ref{eq:l2_penalized} is proportional to the log-posterior for a Bayesian model with Gaussian errors and Gaussian priors on the scores and loadings: 
\begin{align}
\begin{split}
\label{eq:likelihood_for_data}
    \X_{\bigcdot} &\mid \mathbf{U}_{\bigcdot}, \mathbf{V}, \mathbf{W}_{\bigcdot}, \mathbf{V}_{\bigcdot} \sim \prod_{s=1}^q \prod_{i=1}^n \prod_{j=1}^{p_s} \hbox{Normal}\left(\X_s[j,i]| \mathbf{U}_s[j,\cdot]\mathbf{V}[i,\cdot]^T + \mathbf{W}_s[j,\cdot] \mathbf{V}_s[i,\cdot]^T, 1 \right) \\
\end{split}
\end{align}
and 
\begin{align}
\begin{split}
    \mathbf{U}_s[j,\cdot] &\sim \hbox{Normal}(\boldsymbol0, \lambda^{-1} \mathbf{I}_{r\times r}) \\
    \mathbf{V}[i,\cdot] &\sim \hbox{Normal}(\boldsymbol0, \lambda^{-1} \mathbf{I}_{r\times r}) \\
    \mathbf{W}_s[j,\cdot] &\sim \hbox{Normal}(\boldsymbol0, \lambda_s^{-1} \mathbf{I}_{r_s\times r_s}) \\
    \mathbf{V}_s[i,\cdot] &\sim \hbox{Normal}(\boldsymbol0, \lambda_s^{-1} \mathbf{I}_{r_s\times r_s}) \\
\end{split}
\end{align}
Prior to model fitting, we row-center the features and scale the sources to have error variance $1$ using the MAD estimator.  Then, the prior variances for $\mathbf{U}_{\bigcdot}, \mathbf{V}, \mathbf{V}_{\bigcdot}, \mathbf{W}_{\bigcdot}$ are fixed at the penalties described in Section~\ref{ss:UNIFAC} so that the posterior mode of the decomposition of $\X_{\bigcdot}$ matches the UNIFAC decomposition.  We then apply the iterative soft singular value thresholding algorithm of UNIFAC to identify the posterior mode and initialize a Gibbs sampling algorithm to sample from the full posterior distributions of $\mathbf{U}_{\bigcdot}$, $\mathbf{V}$, $\mathbf{W}_{\bigcdot}$, and $\mathbf{V}_{\bigcdot}$.

The factorization for each low-rank term in the decomposition (e.g, $\mathbf{U}_{\bigcdot} \mathbf{V}^T)$ corresponds to a PMF model with $\sigma^2_U=\sigma^2_V$, but the connection to UNIFAC provides several advantages.   First, note that equality of variances does not restrict the model, as their scales are not independently identifiable (see Section~\ref{ss:identifiability}). Moreover, these tuning parameters and the ranks are conveniently fixed via the random matrix theory. As described in \cite{salakhutdinov2008bayesian}, the choice of tuning parameters can dramatically impact model results, and using cross validation with multiple sources of data is not straightforward \citep{owen_bi-cross-validation_2009}. Lastly, an efficient singular value thresholding algorithm can be used to find the mode, circumventing issues of convergence due to poor initialization in Gibbs sampling.      

\subsection{Bayesian Simultaneous Factorization and Prediction (BSFP)}\label{ss:Proposed_Model_y}
Now, suppose we have a continuous phenotype $\y$ in addition to $\X_1, \dots, \X_q$ for a shared cohort of $n$ samples. We are interested in predicting $\y$ using the $q$ sources. We extend the BSF model to include prediction of $\y$ (referred to as BSFP) by assuming the following relationship between the factors, $\mathbf{V}$ and $\mathbf{V}_s$ for $s=1,\dots, q$, and $\y$:
\begin{align}
\begin{split}
\label{eq:likelihood_for_y}
    \y &= \mathbf{V}^* \boldsymbol\beta_{\bigcdot} + \mathbf{e}_y = \beta_0 + \mathbf{V}\boldsymbol\beta_{joint} + \sum_{s=1}^q \mathbf{V}_s \boldsymbol\beta_{indiv,s} + \mathbf{e}_y \\
    \mathbf{e}_y &\sim \hbox{Normal}(\boldsymbol{0}, \tau^2 \mathbf{I}_{n\times n})
\end{split}
\end{align}
where $\mathbf{V}^* = \begin{pmatrix} \boldsymbol{1}_n & \mathbf{V} & \mathbf{V}_1 & \dots & \mathbf{V}_q \end{pmatrix}$ and $\boldsymbol\beta_{\bigcdot} = \begin{pmatrix} \beta_0 & \boldsymbol \beta_{joint}^T & \boldsymbol\beta_{indiv,1}^T & \dots & \boldsymbol\beta_{indiv,q} ^T\end{pmatrix}^T$. 
We use the following priors in our model for $\y$:
\begin{align}
\begin{split}
    \boldsymbol\beta_{\bigcdot} &\sim \hbox{Normal}\left(\boldsymbol0, \boldsymbol{\Sigma}_{\beta} \right) \\
    \tau^2 &\sim \hbox{Inverse-Gamma}(a,b)
\end{split}
\end{align}
where $\boldsymbol\Sigma_{\beta} = diag\{\alpha_0^2, \alpha^2\mathbf{I}_{r\times r}, \alpha^2\mathbf{I}_{r_1 \times r_1}, \dots, \alpha^2\mathbf{I}_{r_q \times r_q} \}$ and the hyperparameters $a$, $b$, $\alpha_0^2$, and $\alpha^2$ are fixed constants. In our study of HIV-associated OLD (Section~\ref{s:data_application}), we fixed $\alpha_0^2 = 1000^2$ and $\alpha^2 = 1$.
This model may be modified to suit the characteristics of a given outcome; for example, we have implemented an analogous model with a probit link for a binary outcome (see Appendix). 
We initialize as in the BSF model, and infer the full posterior for  $\mathbf{U}_{\bigcdot}$, $\mathbf{V}$, $\mathbf{W}_{\bigcdot}$,  $\mathbf{V}_{\bigcdot}$, $\tau^2$, and $\boldsymbol \beta_{\bigcdot}$ via Gibbs sampling, which we describe in more detail in the next section. Inferring the factorization and prediction model simultaneously confers two advantages. First, it incorporates supervision by $\y$ for the latent structures, yielding phenotypically-relevant latent factors. Second, posterior uncertainty in these underlying factors is propagated through to the predictive model. 

We treat $\y$ distinctly from the other $q$ sources because as a vector it does not have low-rank structure. We estimate the error variance in $\y$, $\tau^2$, explicitly, while we fix the error variance in $\X_{\bigcdot}$ to 1 after scaling. Simulations showed that initializing with $\y$ and scaling $\y$ by its estimated error standard deviation, as is done with each $\X_s$, yielded no improvement in recovery of the underlying structure or in prediction (see Appendix for more information). 

\subsection{Model Estimation}\label{ss:estimation}

In this section, we describe the Gibbs algorithm sampling to fit the BSFP model. We assume the sources have been scaled to error variance $1$, as described in Section~\ref{ss:UNIFAC}, and the prior variances on factorization components have been fixed as described in Section~\ref{ss:Proposed_Model}. The steps to sample from the posterior distributions of model parameters for
%
BSFP are as follows:
\begin{enumerate}
    \item Initialize $\mathbf{V}^{(0)}$, $\mathbf{U}^{(0)}_s$, $\mathbf{V}^{(0)}_s$, and $\mathbf{W}^{(0)}_s$ for $s=1,\dots, q$ at the solution to Equation~\ref{eq:nn_obj}. Initialize $\boldsymbol{\beta}_{\bigcdot}^{(0)}$ and $\tau^{2(0)}$ by simulating from their respective prior distributions.
    \item For $t=1,\dots, T$:
    \begin{itemize}
        \item Draw $\mathbf{V}^{(t)}$ from $\mathbf{V}\mid\mathbf{X}_{\bigcdot},\mathbf{y},\mathbf{U}^{(t-1)}_{\bigcdot}, \mathbf{V}^{(t-1)}_{\bigcdot}, \mathbf{W}^{(t-1)}_{\bigcdot}, \boldsymbol{\beta}_{\bigcdot}^{(t-1)}, \tau^{2(t-1)}$. 
        \item Draw $\mathbf{U}^{(t)}_s$ from $\mathbf{U}_s\mid\mathbf{X}_{\bigcdot},\mathbf{y},\mathbf{V}^{(t)}, \mathbf{V}^{(t-1)}_{\bigcdot}, \mathbf{W}^{(t-1)}_{\bigcdot}, \boldsymbol{\beta}_{\bigcdot}^{(t-1)}, \tau^{2(t-1)}$ for $s=1,\dots, q$.
        \item Draw $\mathbf{V}^{(t)}_s$ from $\mathbf{V}_s\mid\mathbf{X}_{\bigcdot},\mathbf{y},\mathbf{U}^{(t)}_{\bigcdot}, \mathbf{V}^{(t)}, \mathbf{W}^{(t-1)}_{\bigcdot}, \boldsymbol{\beta}_{\bigcdot}^{(t-1)}, \tau^{2(t-1)}$ for $s=1,\dots, q$
        \item Draw $\mathbf{W}^{(t)}_s$ from $\mathbf{W}_s\mid\mathbf{X}_{\bigcdot},\mathbf{y},\mathbf{U}^{(t)}_{\bigcdot}, \mathbf{V}^{(t)}, \mathbf{V}^{(t)}_{\bigcdot}, \boldsymbol{\beta}_{\bigcdot}^{(t-1)}, \tau^{2(t-1)}$ for $s=1,\dots, q$.
        \item Draw $\boldsymbol{\beta}^{(t)}_{\bigcdot}$ from $\boldsymbol{\beta}_{\bigcdot}\mid\mathbf{X}_{\bigcdot},\mathbf{y},\mathbf{U}^{(t)}_{\bigcdot}, \mathbf{V}^{(t)}, \mathbf{V}^{(t)}_{\bigcdot}, \tau^{2(t-1)}$.
        \item Draw $\tau^{2(t)}$ from $\tau^2\mid\mathbf{X}_{\bigcdot},\mathbf{y},\mathbf{U}^{(t)}_{\bigcdot}, \mathbf{V}^{(t)}, \mathbf{V}^{(t)}_{\bigcdot}, \boldsymbol{\beta}^{(t)}_{\bigcdot}$.
    \end{itemize}
\end{enumerate}
The Gibbs sampling algorithm for the BSF model is similar, without sampling for or conditioning on $\y$, $\boldsymbol{\beta}_{\bigcdot}$ or $\tau^2$.  Both algorithms are given in the Appendix, with derivations for all full conditional distributions.   In the Appendix we also describe simulations to validate our model fitting algorithm.  

We recommend computing the log-joint density of the model at each sampling iteration and observing trace plots of the estimated structures to assess convergence. Since BSF is initialized at the posterior mode, and BSFP is initialized at the mode conditional on $\X_{\bigcdot}$, a large burn-in is generally not necessary.

\subsection{Identifiability}\label{ss:identifiability}

The joint and individual structures $\mathbf{J}_{\bigcdot}$ and $\mathbf{A}_{\bigcdot}$ in the 
posterior mode of the decomposition of $\X_{\bigcdot}$, which matches the solution to the nuclear-norm penalized objective in Equation~\ref{eq:nn_obj}, are uniquely identified by Theorem 1 of \cite{lock2022bidimensional}. This theorem provides sufficient conditions for identifiability: the decomposition must minimize Equation~\ref{eq:nn_obj} with $\lambda$ and $\lambda_s$ fixed as in Section~\ref{ss:UNIFAC}, and for each source, $s=1,\dots, q$, the columns of the loadings and scores in the joint and individual structures must be linearly independent. 
While the posterior mode is uniquely identified, the decomposition at each posterior sampling iteration may not be as the sampler explores the entire stationary distribution.

A challenge in Bayesian factor models like ours is that the loadings and scores are not identifiable due to rotation, permutation, and sign invariance. Under rotation invariance, $\mathbf{U}_{\bigcdot}\mathbf{V}^T$, for example, is unchanged if we right-multiply $\mathbf{U}_{\bigcdot}$ and $\mathbf{V}$ by an orthogonal $r \times r$ matrix $\mathbf{P}$, i.e. $\mathbf{U}_{\bigcdot}\mathbf{P}(\mathbf{V}\mathbf{P})^T = \mathbf{U}_{\bigcdot}\mathbf{P}\mathbf{P}^T\mathbf{V}^T = \mathbf{U}_{\bigcdot}\mathbf{V}^T$. Under permutation invariance, the columns in $\mathbf{U}_{\bigcdot}$ and $\mathbf{V}$ can be reordered and yield the same decomposition. Likewise, under sign invariance, the signs in the columns of $\mathbf{U}_{\bigcdot}$ and $\mathbf{V}$ can be swapped. Rotation, permutation, and sign invariance obstruct interpretation of posterior summaries of the Gibbs samples for $\mathbf{U}_{\bigcdot}, \mathbf{V}, \mathbf{W}_{\bigcdot}, \mathbf{V}_{\bigcdot}$, and $\boldsymbol{\beta}_{\bigcdot}$. To address all three sources of non-identifiability, we use the MatchAlign algorithm \citep{poworoznek_efficiently_2021} which first orthogonalizes the loadings to load the observed features onto one or a few factors. Then, using a greedy matching algorithm and pivot (described below), factors are iteratively matched to the positively- or negatively-signed pivot columns for which the $L_2$-normed difference is minimized. 

We now describe our adaptation of the MatchAlign algorithm. At each Gibbs sampling iteration after burn-in, $t=T_{\text{burn-in}},\dots, T$, define $\mathbf{U}_{\bigcdot, \beta}^{(t)} = \begin{pmatrix} \mathbf{U}_{\bigcdot}^{(t) T} & \boldsymbol{\beta}_{joint}^{(t)} \end{pmatrix}^T$ and $\mathbf{W}_{s, \beta}^{(t)} = \begin{pmatrix}\mathbf{W}_{s}^{(t) T} & \boldsymbol\beta_{indiv,s}^{(t)} \end{pmatrix}^T$ for $s=1,\dots, q$. 
For each $t$, we apply a Varimax rotation, yielding $\mathbf{U}_{\bigcdot, \beta}^{\dagger(t)}$ and $\mathbf{W}_{s, \beta}^{\dagger(t)}$ for $s=1,\dots, q$. The resulting rotation is also applied to the scores, yielding $\mathbf{V}^{\dagger(t)}$ and $\mathbf{V}_{s}^{\dagger(t)}$ for $s=1,\dots, q$. For pivot, we use $\mathbf{U}_{\bigcdot,\beta}^{\dagger(0)} = \begin{pmatrix} \mathbf{U}_{\bigcdot}^{\dagger(0)T} & \boldsymbol\beta^{\dagger(0)}_{joint} \end{pmatrix}^T$ and $\mathbf{W}_{s,\beta}^{\dagger(0)} = \begin{pmatrix} \mathbf{W}_{s}^{\dagger(0)T} & \boldsymbol\beta^{\dagger(0)}_{indiv,s} \end{pmatrix}^T$, where $\mathbf{U}_{\bigcdot}^{\dagger(0)}$, $\boldsymbol\beta^{\dagger(0)}_{joint}$, $\mathbf{W}_{s}^{\dagger(0)}$, and $ \boldsymbol\beta^{\dagger(0)}_{indiv,s}$ are chosen from the posterior sample with the median condition number \citep{poworoznek_efficiently_2021} after burn-in.
Taking $\mathbf{U}_{\bigcdot, \beta}^{\dagger(t)}$ as an example, we start with the column with the largest norm and calculate the normed difference between each column in the pivot under positive and negative signage, $\mathbf{U}_{\bigcdot,\beta}^{\dagger(0)}[\cdot,k]$ and $-\mathbf{U}_{\bigcdot,\beta}^{\dagger(0)}[\cdot,k]$, $k=1,\dots, r$. We match this column in $\mathbf{U}_{\bigcdot, \beta}^{\dagger(t)}$ to signed $\mathbf{U}_{\bigcdot,\beta}^{\dagger(0)}[\cdot,k]$ with $k$ that yields the smallest norm. We proceed similarly with $\mathbf{W}_{s, \beta}^{\dagger(t)}$ for $s=1,\dots, q$. The scores are permuted and signed appropriately to match. 

\subsection{Multiple Imputation}\label{ss:imputation}

The Gibbs sampling algorithm for estimating the posterior of the latent variation structure and regression coefficients in BSFP naturally accommodates multiple imputation of missing values in $\X_{\bigcdot}$ and $\y$. We use the most up-to-date posterior samples to impute missing values at each iteration of the sampler, even if an entire sample is unobserved in a source (referred to as ``blockwise" missingness). 
Imputations are generated from the posterior predictive distribution and can be studied using standard posterior summaries. 

Let $\mathcal{I}^{(m)}_s = \{(j,i): \X_{s}[j,i] \text{ missing}\}$ denote the set of bivariate indices for which entries in the source $\X_s$ are not observed. At the $t$th iteration of the Gibbs sampler, we impute these entries by randomly generating for $(j,i) \in \mathcal{I}^{(m)}_s$:
\begin{align}
\begin{split}
    \X_{s}[j,i] &= \mathbf{U}_s^{(t)}[j,\cdot] \mathbf{V}[i,\cdot]^{(t)T} + \mathbf{W}_s^{(t)}[j,\cdot] \mathbf{V}[i,\cdot]_s^{(t)T} + \mathbf{E}_s[j,i] 
\end{split}
\end{align}
where $\mathbf{E}_s[j,i] \sim \hbox{Normal}(0, 1)$. Missing values in $\boldsymbol{y}$ may be imputed in a similar manner. Let $\mathcal{I}^{(m)}_{y} = \{i: y_i \text{ missing}\}$. For $i\in \mathcal{I}^{(m)}_{y}$, we may impute these values at iteration $t$ using:
\begin{align}
\begin{split}
    y_i &= \mathbf{V}^{*(t)}[i, \cdot] \boldsymbol{\beta}_{\bigcdot}^{(t)} + e_i 
\end{split}
\end{align}
where $e_i \sim \hbox{Normal}(0, \tau^{2(t)})$.

\section{Simulations}\label{s:Simulation}

We consider two simulation studies to characterize the performance of BSF and BSFP. In Section~\ref{ss:model_comparison}, we compare BSFP to existing one- and two-step approaches to factorizing variation and performing prediction. In Section~\ref{ss:imputation_simulation}, we compare BSF to existing single-imputation approaches using simulated multi-source datasets. 

\subsection{Model Comparison}\label{ss:model_comparison}

We studied via simulation the ability of BSFP to recover latent variation structure and predict a continuous outcome under varying levels of signal-to-noise (s2n). We generated $q=2$ sources of data with $100$ features each on $n=200$ samples, which were then split into a training and test set, $100$ samples apiece, denoted $\X_{\bigcdot}^{train}$, $\y^{train}$, $\X_{\bigcdot}^{test}$, and $\y^{test}$. The true overall rank of the latent structure was $3$, where $r = 1$ and $r_s=1$ for $s=1,2$. We generated $\X_{\bigcdot}$ according to the decomposition in Equation~\ref{eq:likelihood_for_data} where the entries of $\mathbf{U}_{\bigcdot}$, $\mathbf{V}$, $\mathbf{V}_{\bigcdot}$, and $\mathbf{W}_{\bigcdot}$ were generated iid from a $\hbox{Normal}(0,1)$ distribution. We generated $\y$ from the model in Equation~\ref{eq:likelihood_for_y} where an intercept was generated from a $\hbox{Normal}(0,10)$ distribution and $\boldsymbol\beta_{joint}$, $\boldsymbol\beta_{indiv,s}$ were generated iid $\hbox{Normal}(0,1)$. Random noise in $\X_{\bigcdot}$ and $\y$ were generated iid $\hbox{Normal}(0,1)$. We scaled $\X_{\bigcdot}$ and $\y$ to have s2n ratios 9, 3, 1, and $\frac{1}{3}$ and considered all $16$ combinations of s2n.

We considered UNIFAC, JIVE \citep{lock2013joint}, MOFA \citep{argelaguet2018multi}, sJIVE \citep{palzer2022sjive} and BIP \citep{chekouo2020bayesian} as alternative methods. UNIFAC estimates the posterior mode of the data-decomposition of our proposed model and with ranks determined as described in Section~\ref{ss:UNIFAC}. JIVE uses a permutation-based approach to decide the ranks of the joint and individual structures (see Supplementary Materials of \cite{lock2013joint} for more details). MOFA imposes view-wise and factor-wise sparsity to estimate the ranks, but does not distinguish between joint and individual factors. To address this, we treated a MOFA-estimated factor as ``joint" if the maximum amount of variation it explained across sources was within twice the smallest amount of variation explained. 
BIP and sJIVE perform simultaneous factorization and prediction and are the most natural comparisons to BSFP. BIP also does not differentiate between joint and individual factors, so we treated factors as ``joint" if the associated marginal posterior probabilities across the sources were all $>0.5$. For UNIFAC, JIVE, and MOFA, which do not directly accommodate prediction, we treated the estimated factors from each as fixed covariates in a Bayesian linear model for $\y$ as described in Section~\ref{ss:Proposed_Model_y}, following the two-step approach described in Section~\ref{s:intro}. UNIFAC was also an important comparison, as it matches the posterior mode of the data-decomposition in BSFP but does not propagate error from the estimated factorization to prediction. 

We compared how well the models recovered the underlying structure in $\X_{\bigcdot}$ used to train the models and how well each predicted a held-out $\y$ on the test data. The proposed model was fit on the full training and test $\X_{\bigcdot}$ with only access to $\y^{train}$. We used UNIFAC, JIVE, and MOFA to estimate the underlying structure on the full $\X_{\bigcdot}$. The scores corresponding to the test set were then used as covariates to predict $\y^{test}$. sJIVE and BIP were trained on $\X_{\bigcdot}^{train}$ and $\y^{train}$, as these methods do not accommodate prediction of unobserved outcomes. We then used the estimated loadings to predict $\y^{test}$ using $\X_{\bigcdot}^{test}$. We ran each model under each s2n combination for 100 replications and compared their recovery of the underlying structure using the relative squared error (RSE):
\begin{align}
\label{eq:relative_mse_sim}
\hbox{RSE}(\mathbf{S}, \mathbf{\hat{S}}^{mod})= \frac{||\mathbf{S} - \mathbf{\hat{S}}^{mod}||_F^2}{||\mathbf{S}||_F^2} 
\end{align}
where $\mathbf{S} \in \{ \mathbf{J}, \mathbf{A}\}$ reflects the true joint or individual structure, respectively, and $\mathbf{\hat{S}}^{mod} \in \{ \mathbf{\hat{J}}^{mod}, \mathbf{\hat{A}}^{mod}\}$ reflects the estimated joint or individual structures from model, $mod$. An RSE close to 0 suggests better performance. With the exception of UNIFAC, we also consider each method with ranks fixed at the truth. We calculate coverage of the truth for BSFP, UNIFAC, JIVE, and MOFA using 95\% credible intervals. We do not calculate coverage for sJIVE and BIP because neither method offers full posterior inference for the predictive model. 

\begin{figure}[H]
    \centering
    \includegraphics[scale=0.3]{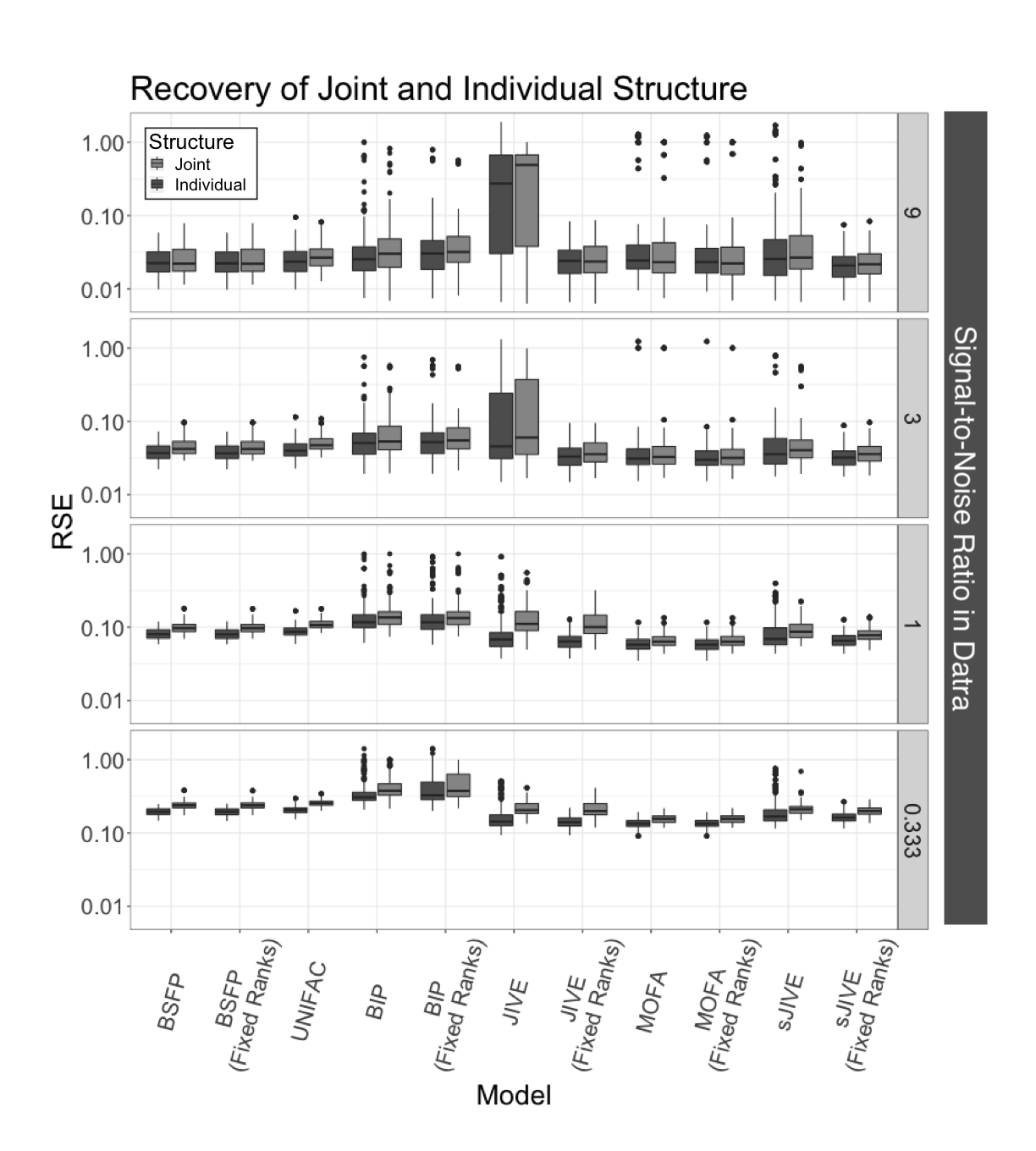}
    \caption{Comparing BSFP to existing methods on how well each recovers underlying joint and individual structure based on relative squared error (RSE). RSE values closer to $0$ reflect better performance. Each panel reflects a different signal-to-noise (s2n) ratio in $\X_{\bigcdot}$. We do not differentiate according to s2n in $\y_{\bigcdot}$ as results did not vary under differing levels of signal in the response. }
    \label{fig:structure_recovery}
\end{figure}

\begin{figure}[H]
    \centering
    \includegraphics[scale=0.27]{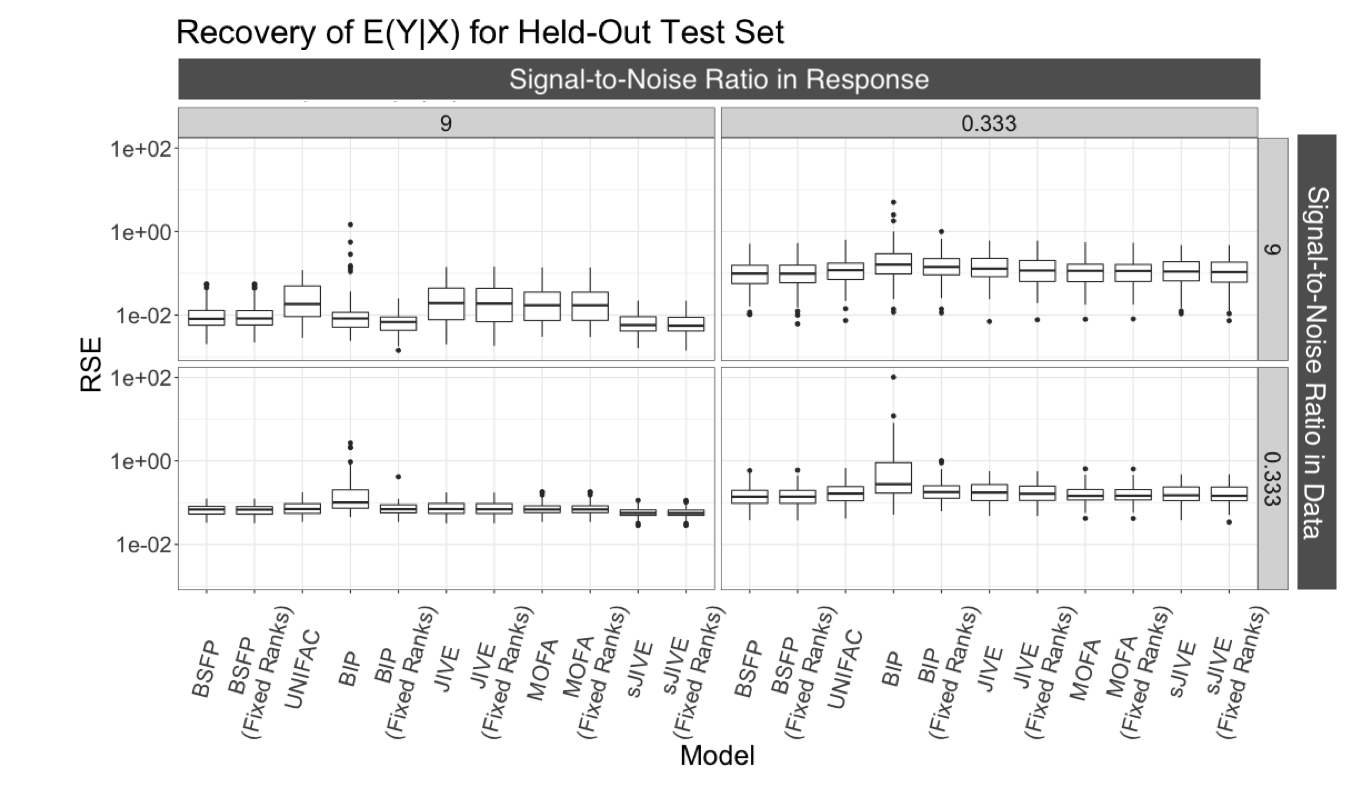}
    \caption{Comparing BSFP to existing methods on how well each recovers the $\mathbb{E}(\y^{test}|\X_{\bigcdot})$ based on relative squared error (RSE). RSE values closer to $0$ reflect better performance. Each panel reflects a different signal-to-noise (s2n) ratio in $\X_{\bigcdot}$ and $\y^{test}$. We select only the highest and lowest s2n ratios for space considerations.}
    \label{fig:ey_test_recovery}
\end{figure}

\begin{figure}[H]
    \centering
    \includegraphics[scale=0.27]{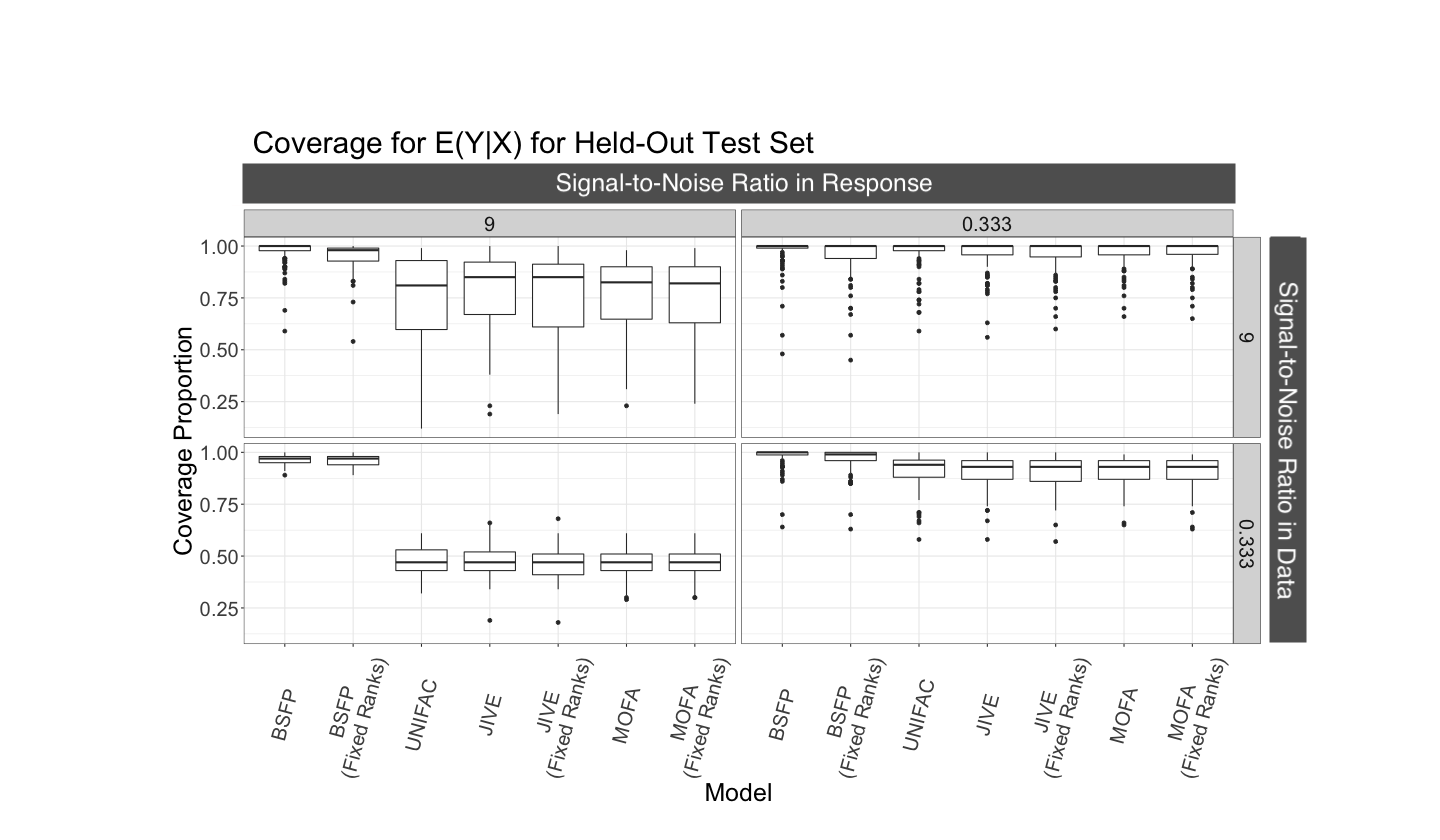}
    \caption{Comparing the proposed model to existing methods on how well each covers the $\mathbb{E}(\y^{test}|\X_{\bigcdot})$ under the posterior. Each panel reflects a different signal-to-noise (s2n) ratio in $\X_{\bigcdot}$ and $\y^{test}$. We select only the highest and lowest s2n ratios for space considerations. Coverage was assessed using 95\% credible intervals. }
    \label{fig:ey_test_coverage}
\end{figure}

The RSEs averaged across simulation replications for recovery of the underlying joint and individual structures is shown in Figure~\ref{fig:structure_recovery} on a log-scale. BSFP, UNIFAC, MOFA, BIP, and sJIVE all performed similarly well in recovering the joint and individual structure, even compared to their performances given the true ranks. With high signal, JIVE overestimated the ranks, leading to poor estimation of the structure. The average relative RSE for recovery of $\mathbb{E}(\y^{test}|\X_{\bigcdot})$ is shown in Figure~\ref{fig:ey_test_recovery} on a log-scale for s2ns $9$ and $1/3$. All models performed comparably well, with BSFP, BIP, and sJIVE showing the smallest median RSE and lowest variability under high signal in $\X_{\bigcdot}$ and $\y$. The greatest benefit of BSFP is apparent when studying coverage of $\mathbb{E}(\y^{test}|\X_{\bigcdot})$, shown in Figure~\ref{fig:ey_test_coverage}. 
BSFP propagates error in estimating the underlying structure to prediction, yielding nominal coverage rates of 95\%. This benefit is especially noticeable when the signal in $\X_{\bigcdot}$ is low and there is more uncertainty in recovering the underlying structure. This suggests BSFP provides reliable inference across varying levels of signal in the data and response.

\subsection{Missing Data Imputation}\label{ss:imputation_simulation}
 
We studied the performance of BSF in imputing missing values. In this simulation, we did not consider prediction of an outcome and focused on how well each method imputes the observations that were randomly removed from each $\X_s$. We generated $q=2$ sources of data, measured on $n=100$ samples, each with $p_s=100$ features, for $s=1,2$. Data were generated in the same manner as described in Section~\ref{ss:model_comparison}.  We considered three different types of missingness: (1) entrywise, in which 10\% of entries in each source were randomly removed, (2) blockwise, in which 10 samples (columns) from each source (non-overlapping) were randomly removed, and (3) missingness-not-at-random (MNAR), in which the lowest 10\% of samples in each source were removed. We varied the s2n in the data across $9$, $3$, $1$, and $1/3$. We considered two different settings for the true ranks: in setting 1, the overall rank was $15$ where $r=r_s=5$ for $s=1,2$, and in setting 2, the overall rank was $3$ as in Section~\ref{ss:model_comparison}. We focus on results from setting 1 and provide a discussion of the results when the overall rank is $3$ in the Appendix. For each condition (missingness type, signal-to-noise level, and overall rank of the underlying structure) we ran each model for $100$ replications and averaged the results across sources and replications. 

We compared BSF to five other single imputation approaches, including mean imputation and UNIFAC. We also compared to the iterative SVD imputation algorithm given by \cite{fuentes2006using} (henceforth referred to simply as SVD) with a rank of $4$, k-nearest neighbors (kNN) imputation \citep{kowarik2016imputation}, and random forest (RF) imputation \citep{stekhoven2012missforest}. We applied the SVD, kNN, and RF to each source independently and to the sources combined. We evaluated the performance of the imputation approaches based on the relative squared error of the imputed values compared to the true unobserved values, i.e. $\hbox{RSE}(\mathbf{X}_s, \hat{\mathbf{X}}^{mod}_s) = ||\{\mathbf{X}_s[j,i]\mid (j,i)\in \mathcal{I}_s^{(m)}\} - \{\hat{\mathbf{X}}^{mod}_s[j,i]\mid (j,i)\in \mathcal{I}_s^{(m)}\}||_F^2/|| \mathbf{X}_s[j,i] \mid (j,i) \in \mathcal{I}_s^{(m)} ||_F^2$ where $\hat{\mathbf{X}}^{mod}_s[j,i]$ denotes the imputation for $\mathbf{X}_s[j,i]$ given by model $mod$. For BSF, the RSE was calculated using the posterior mean of the imputed values across Gibbs sampling iterations. In the Appendix, we describe using BSF for multiple imputation and uncertainty under different types of missingness and different levels of signal.

\begin{figure}[H]
    \centering
    \includegraphics[scale=0.2]{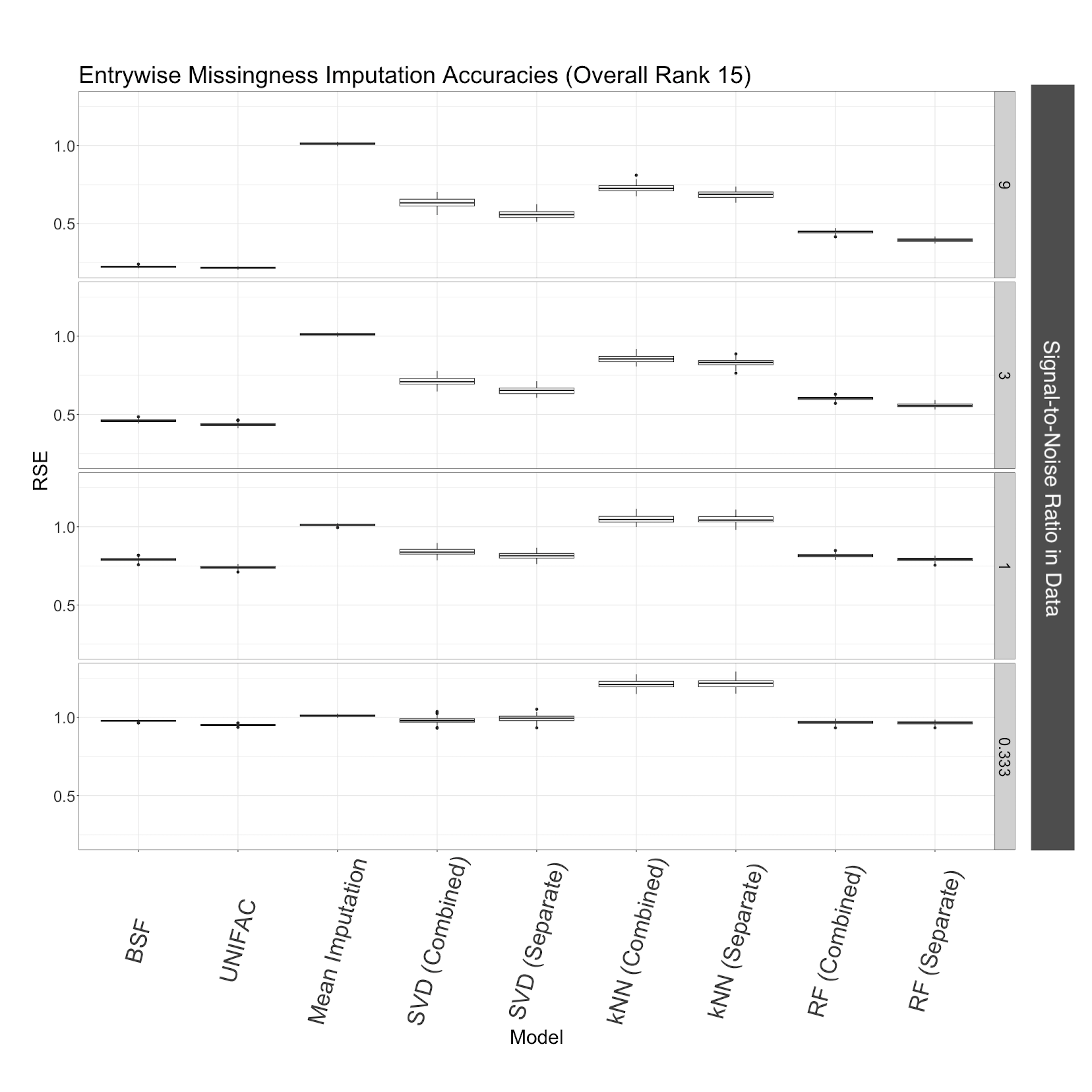}
    \caption{Imputation accuracy results under an assumed overall rank of $15$ under entrywise missingness. Models are compared based on the relative squared error (RSE) of the unobserved values compared to the imputed values. RSE values close to $0$ reflect better performance. For BSF, the RSE is calculated using the posterior mean of the imputed values.}
    \label{fig:entrywise_imputation_rank15}
\end{figure}

Imputation accuracy results for entrywise missingness are shown in Figure~\ref{fig:entrywise_imputation_rank15}. Under high signal ($\hbox{s2n}=9$ and $\hbox{s2n}=3$), both BSF and UNIFAC excel at predicting the unobserved values, with RF closely following in terms of performance. As the signal in the data sources decreases, these differences subside and all methods struggle to accurately impute. 

\begin{figure}[H]
    \centering
    \includegraphics[scale=0.2]{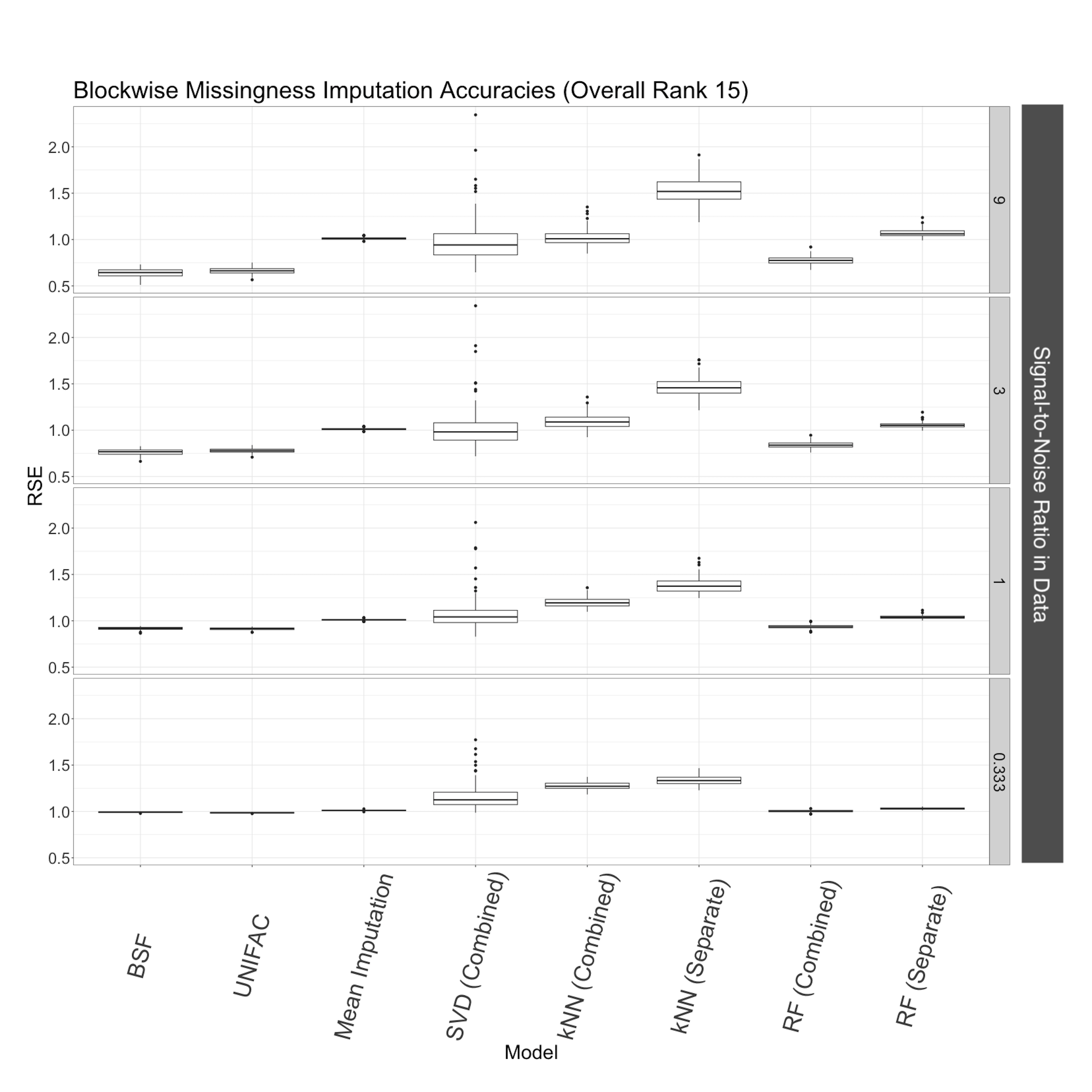}
    \caption{Imputation accuracy results under an assumed overall rank of $15$ under blockwise missingness. Models are compared based on the relative squared error (RSE) of the unobserved values compared to the imputed values. RSE values close to $0$ reflect better performance. For BSF, the RSE is calculated using the posterior mean of the imputed values.}
    \label{fig:blockwise_imputation_rank15}
\end{figure}

Under blockwise missingness, methods that do not estimate an underlying factorization and methods applied to each source independently are not expected to perform well, as there is no information available to predict the missing samples. In fact, the SVD algorithm would not run under this condition and is excluded from the results in Figure~\ref{fig:blockwise_imputation_rank15}. Under high signal, BSF, UNIFAC, and RF applied to the sources combined were the only methods yielding RSEs below $1$ as these methods were able to impute values other than $0$. As in the entrywise missingness case, these gains in performance dissipated as the signal in the data decreased. 

\begin{figure}[H]
    \centering
    \includegraphics[scale=0.2]{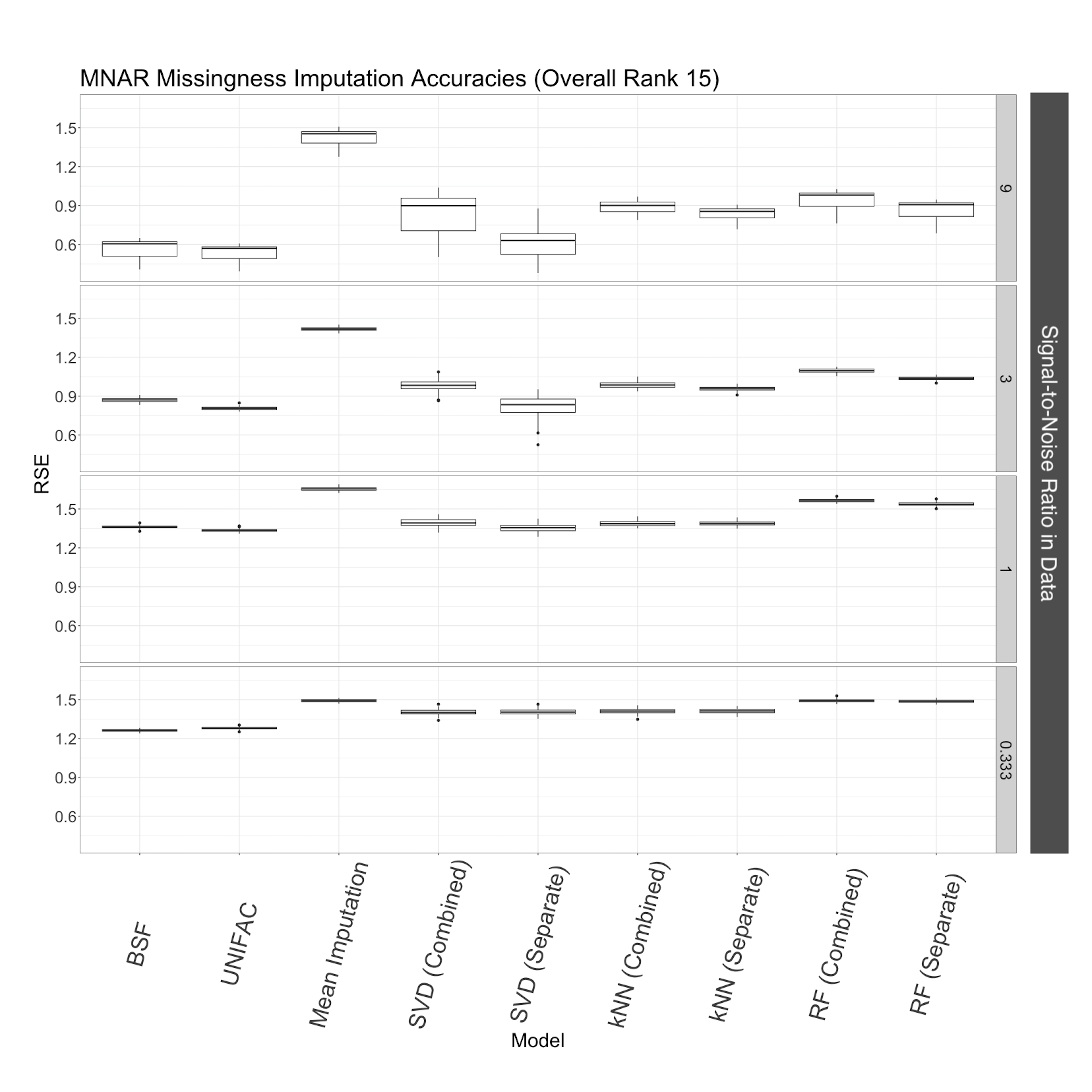}
    \caption{Imputation accuracy results under an assumed overall rank of $15$ under MNAR missingness. Models are compared based on the relative squared error (RSE) of the unobserved values compared to the imputed values. RSE values close to $0$ reflect better performance. For BSF, the RSE is calculated using the posterior mean of the imputed values.}
    \label{fig:MNAR_imputation_rank15}
\end{figure}

Lastly, we considered MNAR (Figure~\ref{fig:MNAR_imputation_rank15}). Under high signal, BSF, UNIFAC, and SVD applied to each source independently provided the best predictive accuracy for unobserved values. This is reasonable given recent results supporting the use of low-rank factorization methods to impute missing values under the MNAR assumption \citep{wang2021matrix}, and underscores the non-viability of methods like kNN and RF in this case. 

\section{Application to HIV-OLD Study}\label{s:data_application}

We applied BSFP to metabolomic and proteomic data collected from the Vancouver and Pittsburgh lung cohorts as part of a matched case-control study on HIV-associated obstructive lung disease (OLD) \citep{cribbs_correlation_2016, akata_altered_2022}. We are interested in predicting lung function based on metabolomic and proteomic expression in bronchoalveolar lavage lung fluid (BALF) among patients living with HIV. In this study, 26 cases (those with OLD) were matched to 26 non-OLD controls based on age, antiretroviral treatment status, and smoking status. Lung function was measured as percent predicted forced expiratory volume in 1-second (FEV1pp). Our dataset contained $252$ BALF metabolites and $4253$ BALF proteins, which we used to predict FEV1pp.

We first compared BSFP to sJIVE, JIVE, UNIFAC, and MOFA under cross validation, where we iteratively held out FEV1pp for each case-control pair. Due to computational barriers, we were unable to fit BIP on these data. We fit BSFP using 5000 posterior samples with a 2500 sample burn-in, which was deemed sufficient based on monitoring of trace plots and the log-joint likelihood of the model. Each model was fit in the same manner as in Section~\ref{s:Simulation}. We also considered the lasso regression model \citep{tibshirani1996regression} fit on the combined metabolomic and proteomic data and fit to each separately. To fit the lasso model, we held-out the FEV1pp values and the metabolomic and/or proteomic observations for each case-control pair and trained the model on the remaining samples. We compared the correlation between the predicted and true held-out FEV1pp values. For BSFP, we used the posterior mean of the predicted FEV1pp values. The correlations and associated p-values given by a Pearson correlation test in descending order were as follows: BSFP (0.4847, $p=0.0003$), UNIFAC (0.4787, $p=0.0003$), sJIVE (0.4425, $p=0.0010$), lasso (combined sources) (0.4029, $p=0.0031$), lasso (metabolite only) (0.3972, $p=0.0035$), MOFA (0.3719, $p=0.0066$), lasso (protein only) (0.3157, $p=0.0226$), and JIVE (0.3131, $p=0.0238$). 
BSFP and UNIFAC had comparable predictive performance, and BSFP has the advantage of providing a framework for uncertainty in the estimated factorization.
In addition, models that distinguished between the joint and individual factors generally performed better, and using one source (only metabolomic or only proteomic) yielded worse results than considering both sources simultaneously. 

\begin{figure}[H]
    \centering
    \includegraphics[scale=0.5]{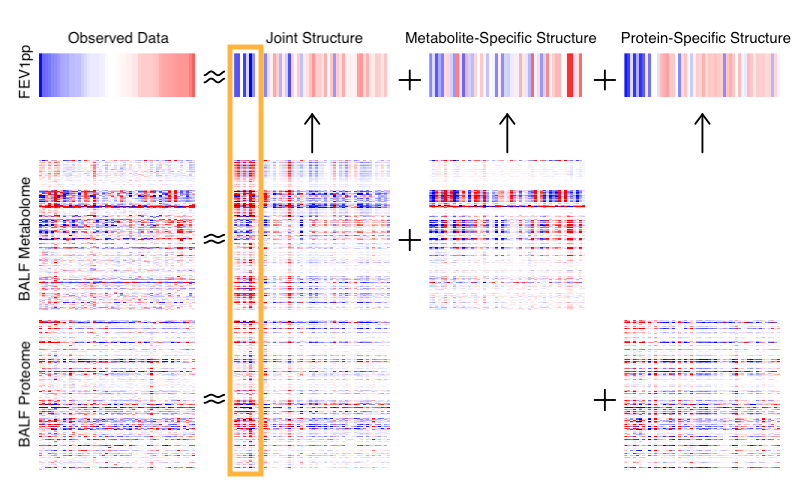}
    \caption{Heatmap of the posterior mean of estimated joint and individual structures using the BALF metabolome and proteome to predict FEV1pp. Columns represent samples and rows represent proteins or metabolites. Samples are ordered by FEV1pp. Blue values reflect lower expression and red values reflect higher expression relative to the rowwise mean. The orange rectangle highlights a cluster of samples with low FEV1pp driven by joint metabolomic and proteomic expression.}
    \label{fig:heatmap}
\end{figure}

\begin{figure}[H]
    \centering
    \includegraphics[scale=0.45]{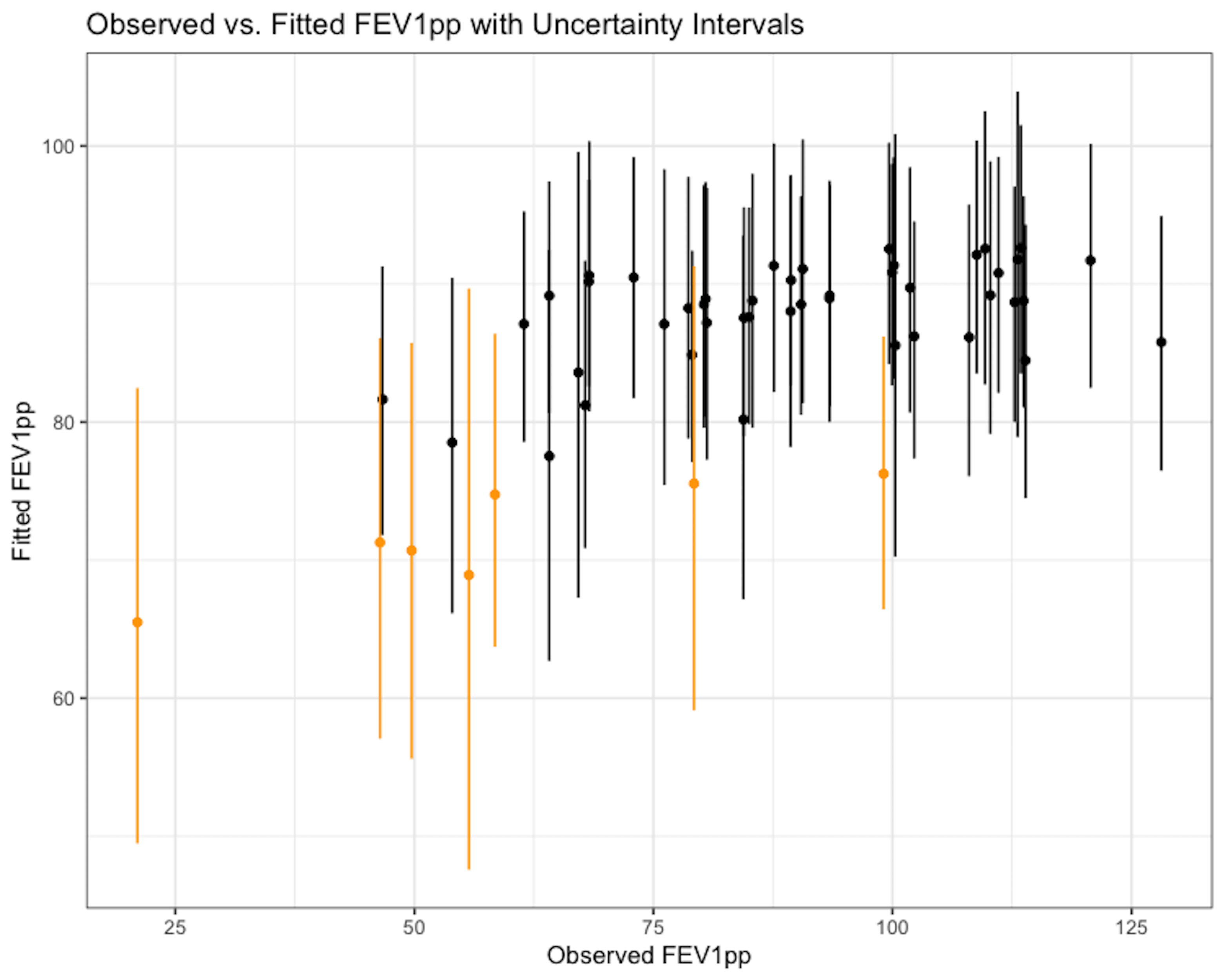}
    \caption{Plot of fitted vs. observed FEV1pp for each of $52$ samples based on BSFP model fit to the full metabolomic and proteomic datasets with 95\% credible intervals. Those colored in orange correspond to the samples which clustered together based on joint metabolomic and proteomic expression. }
    \label{fig:prediction_intervals}
\end{figure}

We now focus on the results from fitting BSFP. BSFP identified $14$ joint, $11$ metabolite-specific, and $16$ protein-specific factors for an overall rank of $41$. We visualize the estimated structures via heatmap (Figure~\ref{fig:heatmap}), where the samples (columns) are ordered by FEV1pp. The joint structure, which explained $17.7$\% of variation in the metabolome (95\% CI=$(16.3\%, 19.2\%)$) and $21.5\%$ in the proteome $(19.5\%, 23.2\%)$, reveals a sample cluster driven by shared metabolomic and proteomic expression, highlighted in orange. 
The individual structures explained $54.6\%$  $(53.5\%, 55.6\%)$ in the metabolome and $61.7\%$ $(60.7\%, 62.9\%)$ in the proteome. The proportion of variance in FEV1pp explained by the joint factors was $2.3$\% $(0.4\%,6.5\%)$, while the metabolomic factors explained $1.2$\% $(0.3\%, 3.0\%)$ and the proteomic factors explained $6.3$\% $(1.5\%, 14.5\%)$. We visualize the accuracy of the fitted FEV1pp vs. observed and the associated uncertainty in fitted FEV1pp in Figure~\ref{fig:prediction_intervals}. There was considerable heterogeneity in the observed FEV1pp, which ranged from 21 to 128\% of predicted normal. The fitted FEV1pp ranged from 73 to 93\% of predicted normal, reflecting the challenge of capturing the full spectrum of FEV1pp heterogeneity in a small sample. Posterior 95\% credible intervals reflect this uncertainty. Samples colored in orange correspond to the samples which clustered together based on shared proteomic and metabolomic expression in Figure~\ref{fig:prediction_intervals}. Clustering was determined using k-means with $k=2$ on the joint structure across the posterior sampling iterations. Samples which clustered together over 90\% of posterior sampling iterations are colored in orange. 
The seven samples in this cluster all had lower fitted FEV1pp than the other 45 samples. Together these results reveal a multi-omic molecular subtype in the lung that is associated with poor lung function; further molecular association with FEV1pp are uncertain.  

\begin{figure}[H]
\centering
\includegraphics[width=0.7\linewidth]{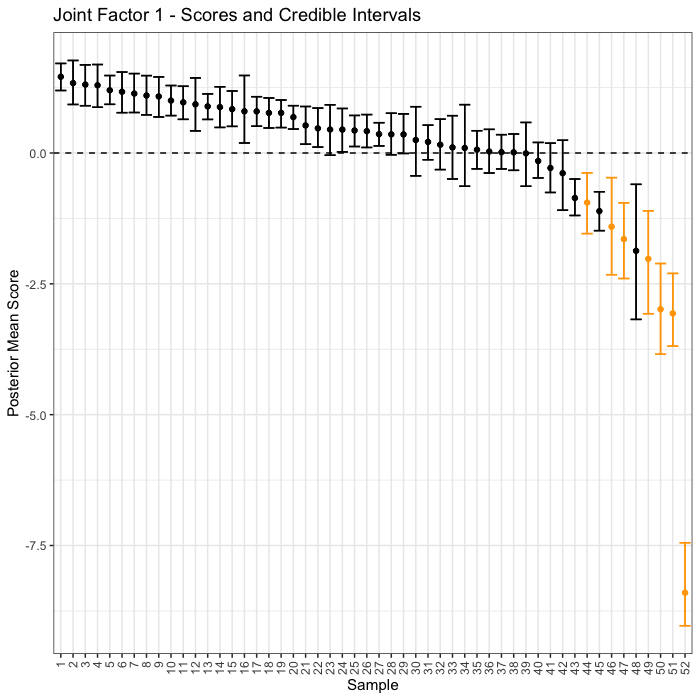}
\caption{Scores for each sample for Joint Factor 1, the joint factor that explains the largest variation within the joint structure. Each point reflects the posterior mean score and the interval reflects the 95\% credible interval. Intervals colored in orange correspond to samples that belong to the stand-out cluster.}
\label{fig:joint_factor_scores}
\end{figure}

\begin{figure}[H]
\centering
\includegraphics[width=0.7\linewidth]{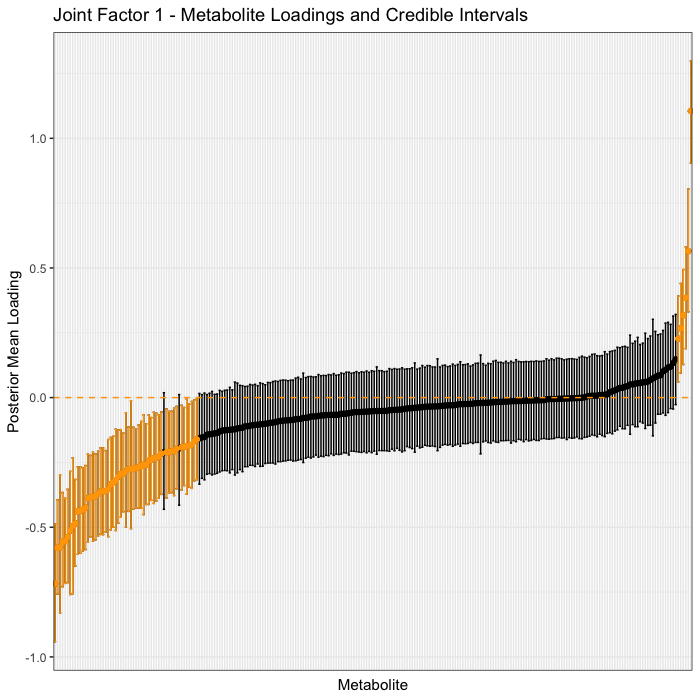}
\caption{Loadings of each observed metabolite for Joint Factor 1. Each point reflects the posterior mean loading and the interval reflects the 95\% credible interval. Intervals colored in orange correspond to those that do not contain 0. }
\label{fig:joint_factor_metab}
\end{figure}

\begin{figure}[H]
\centering
\includegraphics[width=0.7\linewidth]{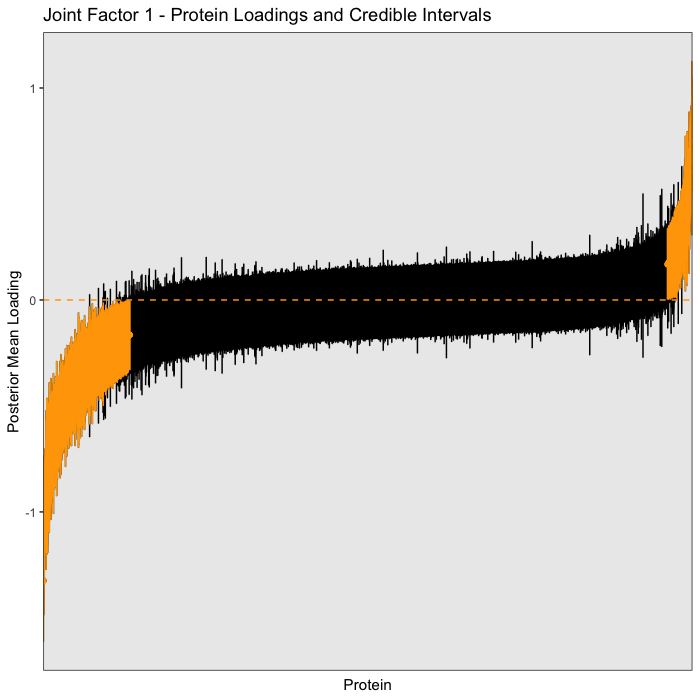}
\caption{Loadings of each observed protein for Joint Factor 1. Each point reflects the posterior mean loading and the interval reflects the 95\% credible interval. Intervals colored in orange correspond to those that do not contain 0. }
\label{fig:joint_factor_prot}
\end{figure}

We applied the alignment algorithm described in Section~\ref{ss:identifiability} to further investigate the factors driving the joint and individual structures. The results were largely unchanged when we considered matching to posterior samples around the chosen pivot. We focus here on the joint factor which explained the largest amount of variation within the joint structure, which we refer to as ``Joint Factor 1," but provide visualizations of all estimated factors in a Shiny app at \url{https://sarahsamorodnitsky.shinyapps.io/BSFP_HIV_OLD/}. 
This component is associated with the previously-identified cluster, as illustrated by the distinct scores for samples within the cluster  (Figure~\ref{fig:joint_factor_scores}). We visualize the loadings of each metabolite and protein in Figures~\ref{fig:joint_factor_metab} and \ref{fig:joint_factor_prot}. Each point in these figures reflects the posterior mean loading of a given biomarker with the associated 95\% credible interval. Intervals colored in orange reflect those that do not contain zero, suggesting these biomarkers contribute ``significantly" to the factor. $61/252$ metabolites and $666/4253$ proteins had ``significant" loadings under the posterior to this joint factor. We used IMPaLA pathway analysis software \citep{kamburov2011integrated} to identify pathways associated with this factor. We treated metabolites and proteins colored in orange as significant and compared against the complete set of metabolites and proteins considered in the analysis as a reference list. The top ten pathways are highlighted in Table~\ref{tab:joint_factor_path}. The top pathways were neutrophil degranulation and innate immunity, both of which are pertinent to OLD. The activation of neutrophils and subsequent inflammation is a hallmark of the disease \citep{herrero2022neutrophils}. In addition to neutrophil-derived inflammation, there is evidence that multiple host defense mechanisms, including innate immunity, play a role in OLD pathogenesis \citep{agusti2019update}.

\begin{table}[H]
\centering
\resizebox{0.9\linewidth}{!}{
\begin{tabular}{ccccccc}
  \hline
Pathway & \begin{tabular}{@{}c@{}} P-Value \\ (Protein)\end{tabular} & \begin{tabular}{@{}c@{}} Q-Value \\ (Protein)\end{tabular} & \begin{tabular}{@{}c@{}} P-Value \\ (Metabolite)\end{tabular} & \begin{tabular}{@{}c@{}} Q-Value \\ (Metabolite)\end{tabular} & \begin{tabular}{@{}c@{}} P-Value \\ (Joint)\end{tabular} & \begin{tabular}{@{}c@{}} Q-Value \\ (Joint)\end{tabular} \\ 
  \hline
Neutrophil degranulation & 0.00000 & 0.00000 & 1.00000 &      1 & 0.00000 & 0.00000 \\ 
  Innate Immune System & 0.00000 & 0.00000 & 0.70600 &      1 & 0.00000 & 0.00000 \\ 
  Immune System & 0.00000 & 0.00000 & 0.92000 &      1 & 0.00000 & 0.00000 \\ 
  Neutrophil extracellular trap formation - Homo sapiens & 0.00000 & 0.00019 & 1.00000 &      1 & 0.00000 & 0.00019 \\ 
  HDMs demethylate histones & 0.00000 & 0.00019 & 1.00000 &      1 & 0.00000 & 0.00019 \\ 
  \begin{tabular}{@{}c@{}}Transcriptional misregulation in cancer \\ - Homo sapiens \end{tabular}
  & 0.00000 & 0.00114 & 1.00000 &      1 & 0.00000 & 0.00114 \\ 
  Complement and coagulation cascades - Homo sapiens  & 0.00000 & 0.00114 & 1.00000 &      1 & 0.00000 & 0.00114 \\ 
  RHO GTPases activate PKNs & 0.00000 & 0.00114 & 1.00000 &      1 & 0.00000 & 0.00114 \\ 
  \begin{tabular}{@{}c@{}}Regulation of Insulin-like Growth Factor (IGF) \\transport and uptake by Insulin-like \\ Growth Factor Binding Proteins (IGFBPs)\end{tabular}
   & 0.00000 & 0.00160 & 1.00000 &      1 & 0.00000 & 0.00160 \\ 
  Signaling by Interleukins & 0.00000 & 0.00201 & 1.00000 &      1 & 0.00000 & 0.00201 \\ 
   \hline
\end{tabular}}
\caption{Top pathways based on the metabolites and proteins with non-zero loadings on ``Joint Factor 1", which explained the most variation across the metabolomic and proteomic datasets. }
\label{tab:joint_factor_path}
\end{table}

Lastly, we studied the imputation accuracy of BSFP on the metabolomic and proteomic data. We considered entrywise and blockwise missingness conditions, where we randomly removed 10\% of observations in a similar manner described in Section~\ref{ss:imputation_simulation}. We compared BSFP to mean imputation, UNIFAC, and SVD, kNN, and RF on the metabolomic and proteomic datasets combined and each source separately. Models were evaluated based on the RSE of their imputations for unobserved values. Table~\ref{tab:imputation_results} summarizes the imputation results averaged across $20$ replications. BSFP and UNIFAC achieved better imputation accuracy than the SVD and mean imputation, and BSFP showed the best performance under blockwise missingness. However, all methods performed poorly under blockwise missingness, and the BSFP results suggest the BALF metabolome cannot be used to accurately impute the proteome, and vice versa. 

\begin{table}[H]
\centering
\caption{Imputation accuracy for randomly-removed values in BALF metabolomic and proteomic data. Each entry reflects the relative square error (RSE) averaged across 20 replications with the standard deviation across replications in parentheses. Under entrywise missingness, $10\%$ of observations were randomly selected across both sources to remove. Under blockwise missingness, $5$ non-overlapping samples were randomly removed from each source. }
\label{tab:imputation_results}
\resizebox{0.9\linewidth}{!}{\begin{tabular}{cccc}
  \hline
Model & Missingness & Mean Metabolome RSE (SD) & Mean Proteome RSE (SD) \\ 
  \hline
  BSFP & Entrywise & 0.629 (0.044) & 0.416 (0.007) \\ 
  UNIFAC & Entrywise & 0.631 (0.04) & 0.405 (0.008) \\ 
  Mean Imputation & Entrywise & 1.042 (0.009) & 1.042 (0.002) \\ 
  SVD (Combined) & Entrywise & 1.078 (0.499) & 0.562 (0.038) \\ 
  SVD (Separate) & Entrywise & 1.421 (0.636) & 0.556 (0.031) \\ 
  kNN (Combined) & Entrywise & 1.239 (0.1) & 0.705 (0.015) \\ 
  kNN (Separate) & Entrywise & 1.058 (0.14) & 0.708 (0.016) \\ 
  RF (Combined) & Entrywise & 0.853 (0.038) & 0.489 (0.008) \\ 
  RF (Separate) & Entrywise & 0.659 (0.05) & 0.487 (0.008) \\ 
   \hline
  BSFP & Blockwise & 0.965 (0.022) & 0.991 (0.004) \\ 
  UNIFAC & Blockwise & 0.999 (0.016) & 1.017 (0.015) \\ 
  Mean Imputation & Blockwise & 1.044 (0.02) & 1.038 (0.02) \\ 
  SVD (Combined) & Blockwise & 1.463 (1.089) & 1.153 (0.31) \\ 
  kNN (Combined) & Blockwise & 1.398 (0.208) & 3.621 (1.356) \\ 
  kNN (Separate) & Blockwise & 1.715 (0.261) & 3.637 (1.341) \\ 
  RF (Combined) & Blockwise & 0.997 (0.061) & 1.186 (0.225) \\ 
  RF (Separate) & Blockwise & 1.137 (0.07) & 1.26 (0.242) \\ 
  \hline
\end{tabular}}
\end{table}

\section{Discussion}\label{s:discussion}

In this article, we proposed two Bayesian approaches for factorization of multi-omic data, BSF and BSFP. BSF decomposes variation across the omics datasets into joint and individual structures and samples from the posterior distributions of these structures, accommodating full posterior inference. BSFP incorporates prediction of a clinical outcome or biological phenotype in a complete framework for uncertainty. This allows for the propagation of error associated with the estimated factorization into the predictive model. We showed via simulation the importance of propagating uncertainty. Otherwise, posterior inference in the predictive model will not be appropriate and may not cover the truth. In cases when the signal in the omics sources is low, it is especially important to account for uncertainty in the estimated factorization when using the estimated factors to predict an outcome or phenotype.

Our simulations on imputation accuracy suggested that BSF can be used under several different missingness mechanisms and patterns. When observations were randomly removed from each source, BSF was competitive against existing methods for imputing the unobserved values. Under blockwise missingness and MNAR, BSF yielded gains in imputation accuracy. In addition, BSF offers full posterior inference for the imputed values, which can be studied using posterior summaries. 

Our data application revealed an interesting cluster of participants with HIV-associated OLD driven by shared metabolomic and proteomic expression patterns. While the results shed light on some disease pathways that may have been disrupted, more research needs to be done to validate if a novel OLD subtype exists based on BALF metabolite and protein expression in other cohorts. While BSFP was able to improve upon existing methods in terms of prediction accuracy, it was challenging to accurately capture the heterogeneity in FEV1pp in this dataset. One challenge may have been the low sample size ($52$) relative to the high rank of the estimated factorization ($41$). In the future, considering a steeper penalty in the initialization of the model (i.e., smaller prior variance on the factorization components) could yield a lower-ranked factorization with more distinguished factors. 

There are several avenues for further development. We assumed the omics sources contained real-valued continuous entries, but this could be generalized to other parametric distributions. Accommodating ``bidimensional" structure, when we have multiple omics sources measured on multiple cohorts of samples, would also be worthwhile. Lastly, accommodating longitudinal data structures may reveal molecular disease processes beyond what is found from a cross-sectional analysis. 




\section*{Acknowledgements}

This work was supported by NIH grants R01-GM130622 and R01-HL140971. The views expressed in this article are those of the authors and do not reflect the views of the United States Government, the Department of Veterans Affairs, the funders, the sponsors, or any of the authors’ affiliated academic institutions.

\section*{Software}

Software to run BSFP is provided in an R package on Github at \url{https://github.com/sarahsamorodnitsky/BSFP}. All analysis code is available on a Github repository, linked \url{https://github.com/sarahsamorodnitsky/BSFP_Analysis}.

\appendix
\section{Gibbs Sampler Details}
\label{gibbs_sampler}

\subsection{Gibbs Sampling Algorithm Steps}
\label{gibbs_sampler_steps}

The steps to sample from the posterior distributions of model parameters for BSF are as follows:
\begin{enumerate}
    \item Initialize $\mathbf{V}^{(0)}$, $\mathbf{U}^{(0)}_s$, $\mathbf{V}^{(0)}_s$, and $\mathbf{W}^{(0)}_s$ for $s=1,\dots, q$ %
    via UNIFAC to give the posterior mode.
    \item For $t=1,\dots, T$:
    \begin{itemize}
        \item Draw $\mathbf{V}^{(t)}$ from $\mathbf{V}\mid\mathbf{X}_{\bigcdot},\mathbf{U}^{(t-1)}_{\bigcdot}, \mathbf{V}^{(t-1)}_{\bigcdot}, \mathbf{W}^{(t-1)}_{\bigcdot}$ as given in Equation~\ref{eq:BSF_con_pos_V}
        \item Draw $\mathbf{U}^{(t)}_s$ from $\mathbf{U}_s\mid\mathbf{X}_{\bigcdot},\mathbf{V}^{(t)}, \mathbf{V}^{(t-1)}_{\bigcdot}, \mathbf{W}^{(t-1)}_{\bigcdot}$ for $s=1,\dots, q$ as given in Equation~\ref{eq:con_pos_U}
        \item Draw $\mathbf{V}^{(t)}_s$ from $\mathbf{V}_s\mid\mathbf{X}_{\bigcdot},\mathbf{U}^{(t)}_{\bigcdot}, \mathbf{V}^{(t)}, \mathbf{W}^{(t-1)}_{\bigcdot}$ for $s=1,\dots, q$ as given in Equation~\ref{eq:BSF_con_pos_Vs}
        \item Draw $\mathbf{W}^{(t)}_s$ from $\mathbf{W}_s\mid\mathbf{X}_{\bigcdot},\mathbf{U}^{(t)}_{\bigcdot}, \mathbf{V}^{(t)}, \mathbf{V}^{(t)}_{\bigcdot}$ for $s=1,\dots, q$ as given in Equation~\ref{eq:con_pos_Ws}
    \end{itemize}
\end{enumerate}

The steps to sample from the posterior distributions of model parameters for BSFP are as follows:
\begin{enumerate}
    \item Initialize $\mathbf{V}^{(0)}$, $\mathbf{U}^{(0)}_s$, $\mathbf{V}^{(0)}_s$, and $\mathbf{W}^{(0)}_s$ for $s=1,\dots, q$ 
    via UNIFAC to give the posterior mode for $\X$. Initialize $\boldsymbol{\beta}_{\bigcdot}^{(0)}$ and $\tau^{2(0)}$ by simulating from their respective prior distributions.
    \item For $t=1,\dots, T$:
    \begin{itemize}
        \item Draw $\mathbf{V}^{(t)}$ from $\mathbf{V}\mid\mathbf{X}_{\bigcdot},\mathbf{y},\mathbf{U}^{(t-1)}_{\bigcdot}, \mathbf{V}^{(t-1)}_{\bigcdot}, \mathbf{W}^{(t-1)}_{\bigcdot}, \boldsymbol{\beta}_{\bigcdot}^{(t-1)}, \tau^{2(t-1)}$ as given in Equation~\ref{eq:BSFP_con_pos_V}
        \item Draw $\mathbf{U}^{(t)}_s$ from $\mathbf{U}_s\mid\mathbf{X}_{\bigcdot},\mathbf{y},\mathbf{V}^{(t)}, \mathbf{V}^{(t-1)}_{\bigcdot}, \mathbf{W}^{(t-1)}_{\bigcdot}, \boldsymbol{\beta}_{\bigcdot}^{(t-1)}, \tau^{2(t-1)}$ for $s=1,\dots, q$ as given in Equation~\ref{eq:con_pos_U}
        \item Draw $\mathbf{V}^{(t)}_s$ from $\mathbf{V}_s\mid\mathbf{X}_{\bigcdot},\mathbf{y},\mathbf{U}^{(t)}_{\bigcdot}, \mathbf{V}^{(t)}, \mathbf{W}^{(t-1)}_{\bigcdot}, \boldsymbol{\beta}_{\bigcdot}^{(t-1)}, \tau^{2(t-1)}$ for $s=1,\dots, q$ as given in Equation~\ref{eq:BSFP_con_pos_Vs}
        \item Draw $\mathbf{W}^{(t)}_s$ from $\mathbf{W}_s\mid\mathbf{X}_{\bigcdot},\mathbf{y},\mathbf{U}^{(t)}_{\bigcdot}, \mathbf{V}^{(t)}, \mathbf{V}^{(t)}_{\bigcdot}, \boldsymbol{\beta}_{\bigcdot}^{(t-1)}, \tau^{2(t-1)}$ for $s=1,\dots, q$ as given in Equation~\ref{eq:con_pos_Ws}
        \item Draw $\boldsymbol{\beta}^{(t)}_{\bigcdot}$ from $\boldsymbol{\beta}_{\bigcdot}\mid\mathbf{X}_{\bigcdot},\mathbf{y},\mathbf{U}^{(t)}_{\bigcdot}, \mathbf{V}^{(t)}, \mathbf{W}^{(t)}_{\bigcdot}, \mathbf{V}^{(t)}_{\bigcdot}, \tau^{2(t-1)}$ as given in Equation~\ref{eq:BSFP_con_pos_beta}
        \item Draw $\tau^{2(t)}$ from $\tau^2\mid\mathbf{X}_{\bigcdot},\mathbf{y},\mathbf{U}^{(t)}_{\bigcdot}, \mathbf{V}^{(t)}, \mathbf{W}^{(t)}_{\bigcdot}, \mathbf{V}^{(t)}_{\bigcdot}, \boldsymbol{\beta}^{(t)}_{\bigcdot}$ as given in Equation~\ref{eq:BSFP_con_pos_tau2}
    \end{itemize}
\end{enumerate}

\subsection{Conditional Posteriors}
\label{conditional_posteriors}

We now describe the conditional posteriors for BSF and BSFP in general terms. In the main manuscript, we described scaling each $\mathbf{X}_s$ to have error variance $1$, so we fix $\sigma^2_s = 1$ for $s=1,\dots, q$. We also motivate fixing $\lambda^{-1} = \left(\sqrt{n} + \sqrt{p} \right)^{-1}$
where $p=\sum_{s=1}^q p_s$ and 
$\lambda_s^{-1} = \left(\sqrt{n}+ \sqrt{p_s} \right)^{-1}$. In the following, we define $\boldsymbol\Sigma_s = \sigma^2_s \mathbf{I}_{p_s \times p_s}$ and $\boldsymbol\Sigma = diag\{\boldsymbol\Sigma_1, \boldsymbol\Sigma_2, \dots, \boldsymbol\Sigma_q \}$.\\

\noindent 
Conditional posterior for $\mathbf{V}$ in BSF: for $i=1,\dots, n$, we sample from the posterior distribution for $\mathbf{V}$ by sampling from 

\begin{equation}\label{eq:BSF_con_pos_V}
P(\mathbf{V}[i,]\mid\mathbf{X}_{\bigcdot}, \mathbf{U}_{\bigcdot}, \mathbf{W}_{\bigcdot}, \mathbf{V}_{\bigcdot}, \{\sigma^2_s\}_{s=1}^q, \lambda^{-1}, \{\lambda^{-1}_{s} \}_{s=1}^q) = \hbox{MVN} \left(\mathbf{B}_V \mathbf{b}_V, \mathbf{B}_V \right) 
\end{equation}
where:
\begin{align*} 
    \mathbf{B}_V^{-1} &= \mathbf{U}_{\bigcdot}^T \mathbf{\Sigma}^{-1} \mathbf{U}_{\bigcdot} + \frac{1}{\lambda^{-1}} \mathbf{I}_{r \times r} \\
    \mathbf{b}_V &= \mathbf{U}_{\bigcdot}^T \mathbf{\Sigma}^{-1} (\mathbf{X}_{\bigcdot} - \mathbf{W}_{\bigcdot} \mathbf{V}_{\bigcdot}^T)[,i]
\end{align*}
and MVN represents the probability density function for the multivariate normal distribution. \\

\noindent
Conditional posterior for $\mathbf{V}$ in BSFP: if a continuous response vector, $\mathbf{y}$, is given, we sample from the posterior distribution for $\mathbf{V}$ by sampling from 

\begin{equation}\label{eq:BSFP_con_pos_V}
P(\mathbf{V}[i,]\mid\mathbf{X}_{\bigcdot}, \mathbf{U}_{\bigcdot}, \mathbf{W}_{\bigcdot}, \mathbf{V}_{\bigcdot}, \{\sigma^2_s\}_{s=1}^q, \lambda^{-1}, \{\lambda^{-1}_{s} \}_{s=1}^q, \mathbf{y}, \boldsymbol{\beta}_{\bigcdot}, \tau^2) = \hbox{MVN} \left(\mathbf{B}_V \mathbf{b}_V, \mathbf{B}_V \right)
\end{equation}
where:
\begin{align*}
    \mathbf{B}_V^{-1} &= \begin{pmatrix} \mathbf{U_{\bigcdot}} \\ \boldsymbol{\beta}_{\,joint}^T \end{pmatrix}^T \begin{pmatrix} \mathbf{\Sigma} & \boldsymbol{0} \\ \boldsymbol{0} & \tau^2 \end{pmatrix}^{-1}\begin{pmatrix} \mathbf{U_{\bigcdot}} \\ \boldsymbol{\beta}_{\,joint}^T \end{pmatrix} + \frac{1}{\lambda^{-1}} \mathbf{I}_{r\times r} \\
    \mathbf{b}_V &= \begin{pmatrix} \mathbf{U_{\bigcdot}} \\ \boldsymbol{\beta}_{\,joint}^T \end{pmatrix}^T \begin{pmatrix} \mathbf{\Sigma} &  \boldsymbol{0}\\ \boldsymbol{0} & \tau^2 \end{pmatrix}^{-1} \begin{pmatrix}
    (\mathbf{X}_{\bigcdot} - \mathbf{W}_{\bigcdot} \mathbf{V}_{\bigcdot}^T)[,i] \\
    (\mathbf{y} -\beta_0 - \sum_{s=1}^q \mathbf{V}_s \boldsymbol{\beta}_{indiv,s})[i,]
    \end{pmatrix}
\end{align*}

\noindent
Conditional Posterior for $\mathbf{U}_{\bigcdot}$ in BSF and BSFP: for $s=1,\dots, q$ and $j = 1,\dots, p_s$, we can sample from the posterior distribution for $\mathbf{U}_s$ by sampling from 

\begin{equation}\label{eq:con_pos_U}
P(\mathbf{U}_s[j,]\mid\mathbf{X}_{\bigcdot}, \mathbf{V}, \mathbf{W}_{\bigcdot}, \mathbf{V}_{\bigcdot}, \{\sigma^2_s\}_{s=1}^q, \lambda^{-1}, \{\lambda^{-1}_{s} \}_{s=1}^q) = \hbox{MVN}(\mathbf{B}_{U_s} \mathbf{b}_{U_s}, \mathbf{B}_{U_s})     
\end{equation}
where:
\begin{align*}
    \mathbf{B}_{U_s}^{-1} &= \frac{1}{\sigma^2_s} \mathbf{V^T} \mathbf{V} + \frac{1}{\lambda^{-1}} \mathbf{I}_{r\times r} \\
    \mathbf{b}_{U_s} &= \frac{1}{\sigma^2_s} \mathbf{V}^T (\mathbf{X}_s - \mathbf{W}_s \mathbf{V}_s^T)[j,]
\end{align*}

\noindent
Conditional Posterior for $\mathbf{V}_s$ for BSF: for $s=1,\dots, q$ and $i=1,\dots, n$, we can sample from the posterior distribution of $\mathbf{V}_s$ by sampling from 

\begin{equation}\label{eq:BSF_con_pos_Vs}
P(\mathbf{V}_s[i,]\mid\mathbf{X}_{\bigcdot}, \mathbf{V}, \mathbf{W}_{\bigcdot}, \mathbf{U}_{\bigcdot}, \{\sigma^2_s\}_{s=1}^q, \lambda^{-1}, \{\lambda^{-1}_{s} \}_{s=1}^q) = \hbox{MVN}(\mathbf{B}_{V_s} \mathbf{b}_{V_s}, \mathbf{B}_{V_s})    
\end{equation}
where:
\begin{align*}
    \mathbf{B}_{V_s}^{-1} &= \mathbf{W}_s^T \mathbf{\Sigma}_s^{-1} \mathbf{W}_s + \frac{1}{\lambda^{-1}_{s}} \mathbf{I}_{r_s\times r_s} \\
    \mathbf{b}_{V_s} &= \mathbf{W}_s^T \mathbf{\Sigma}^{-1}_s (\mathbf{X}_s - \mathbf{U}_s \mathbf{V}^T)[,i]
\end{align*}

\noindent
Conditional Posterior for $\mathbf{V}_s$ for BSFP: If a continuous response vector, $\mathbf{y}$, is given, for $s=1,\dots,q$ and $i=1,\dots,n$ we can sample from the posterior distribution of $\mathbf{V}_s$ by sampling from:

\begin{equation}\label{eq:BSFP_con_pos_Vs}
P(\mathbf{V}_s[i,]\mid\mathbf{X}_{\bigcdot}, \mathbf{V}, \mathbf{W}_{\bigcdot}, \mathbf{U}_{\bigcdot}, \{\sigma^2_s\}_{s=1}^q, \lambda^{-1}, \{\lambda^{-1}_{s} \}_{s=1}^q, \mathbf{y}, \boldsymbol{\beta}_{\bigcdot}, \tau^2) = \hbox{MVN}(\mathbf{B}_{V_s} \mathbf{b}_{V_s}, \mathbf{B}_{V_s}) \\    
\end{equation}
where:
\begin{align*}
    \mathbf{B}_{V_s}^{-1} = \begin{pmatrix}
    \mathbf{W}_s \\
    \boldsymbol{\beta}_{indiv,s}
    \end{pmatrix}^T \begin{pmatrix}
    \mathbf{\Sigma}_s & \boldsymbol{0} \\
    \boldsymbol{0} & \tau^2
    \end{pmatrix}^{-1}\begin{pmatrix}
    \mathbf{W}_s \\
    \boldsymbol{\beta}_{indiv,s}
    \end{pmatrix} + \frac{1}{\lambda^{-1}_{s}} \mathbf{I}_{r_s\times r_s} \\
    \mathbf{b}_{V_s} = \begin{pmatrix}
    \mathbf{W}_s \\
    \boldsymbol{\beta}_{indiv,s}
    \end{pmatrix}^T \begin{pmatrix}
    \mathbf{\Sigma}_s & \boldsymbol{0} \\
    \boldsymbol{0} & \tau^2
    \end{pmatrix}^{-1} \begin{pmatrix}
    (\mathbf{X}_s - \mathbf{U}_s \mathbf{V}^T)[,i] \\
    (\mathbf{y} - \beta_0 - \mathbf{V} \boldsymbol{\beta}_{joint} - \sum_{s' \neq s} \mathbf{V}_{s'} \boldsymbol{\beta}_{indiv,s'})[,i]
    \end{pmatrix}
\end{align*} 

\noindent
Conditional Posterior for $\mathbf{W}_{\bigcdot}$ for BSF and BSFP: for $s=1,\dots, q$ and $j=1,\dots, p_s$, we can sample from the posterior distribution for $\mathbf{W}_s$ by sampling from 

\begin{equation}\label{eq:con_pos_Ws}
P(\mathbf{W}_s[j,]\mid\mathbf{X}_{\bigcdot}, \mathbf{V}, \mathbf{V}_{\bigcdot}, \mathbf{U}_{\bigcdot}, \{\sigma^2_s\}_{s=1}^q, \lambda^{-1}, \{\lambda^{-1}_{s} \}_{s=1}^q) = \hbox{MVN}(\mathbf{B}_{W_s} \mathbf{b}_{W_s}, \mathbf{B}_{W_s})
\end{equation}
where:
\begin{align}
    \mathbf{B}_{W_s}^{-1} &= \frac{1}{\sigma^2_s} \mathbf{V}_s^T  \mathbf{V}_s + \frac{1}{\lambda^{-1}_{s}} \mathbf{I}_{r_s\times r_s} \\
    \mathbf{b}_{W_s} &= \frac{1}{\sigma^2_s} \mathbf{V}_s^T  (\mathbf{X}_s - \mathbf{U}_s \mathbf{V}^T)[j,]
\end{align}

\noindent
Conditional Posterior for $\boldsymbol{\beta}_{\bigcdot}$ for BSFP: if $\mathbf{y}$ is continuous, we can sample from the posterior for $\boldsymbol{\beta}_{\bigcdot}$ by sampling from:

\begin{equation}\label{eq:BSFP_con_pos_beta}
P(\boldsymbol{\beta}_{\bigcdot}\mid\mathbf{X}_{\bigcdot}, \mathbf{V}, \mathbf{V}_{\bigcdot}, \mathbf{U}_{\bigcdot}, \{\sigma^2_s\}_{s=1}^q, \lambda^{-1}, \{\lambda^{-1}_{s} \}_{s=1}^q, \mathbf{y}, \tau^2) = \hbox{MVN}(\mathbf{B}_\beta \mathbf{b}_\beta, \mathbf{B}_\beta) 
\end{equation}
where:
\begin{align}
    \mathbf{B}_\beta^{-1} &= \frac{1}{\tau^2} \mathbf{V}^{*T} \mathbf{V}^{*} + \mathbf{\Sigma}_\beta^{-1} \\
    \mathbf{b}_\beta &= \frac{1}{\tau^2} \mathbf{V}^{*T} \mathbf{y}
\end{align}
where $\mathbf{V}^{*} = \begin{pmatrix} \mathbf{1} & \mathbf{V} & \mathbf{V}_1 & \dots & \mathbf{V}_q \end{pmatrix}$.

\noindent
Conditional Posterior for $\tau^2$ for BSFP: if $\mathbf{y}$ is continuous, we can sample from the posterior for $\tau^2$ by sampling from:
\begin{equation}\label{eq:BSFP_con_pos_tau2}
    P(\tau^2\mid\mathbf{X}_{\bigcdot}, \mathbf{V}, \mathbf{V}_{\bigcdot}, \mathbf{U}_{\bigcdot}, \{\sigma^2_s\}_{s=1}^q, \lambda^{-1}, \{\lambda^{-1}_{s} \}_{s=1}^q, \mathbf{y}, \boldsymbol{\beta}_{\bigcdot}) = \hbox{Inverse-Gamma}\left(a + \frac{n}{2}, b + \frac{1}{2} \sum_{i=1}^n (y_i - \mathbf{V}^*[i,]\boldsymbol{\beta}_{\bigcdot})^2 \right)
\end{equation}

\section{BSFP with Binary Outcome}\label{sup:binary}

\subsection{Model Formulation}

We describe formulating our model assuming the outcome variable is binary. We can model the outcome by assuming $y_i | \mathbf{V}, \mathbf{V}_{\bigcdot}, \boldsymbol{\beta}_{\bigcdot} \sim \hbox{Bernoulli}(\Phi(\mathbf{V}^*\boldsymbol{\beta}))$ where $\Phi(\cdot)$ represents the cumulative distribution function of a standard normal distribution. As discussed in \cite{albert1993bayesian}, we introduce a latent variable $z_i$ for every $y_i$ to facilitate Gibbs sampling. We define $z_i$ such that $y_i = 1$ if $z_i >0$ and $y_i = 0$ if $z_i < 0$. The full model framework with priors is defined as follows:

\begin{align}
\begin{split}
    \mathbf{y}|\X_{\bigcdot}, \mathbf{V}, \mathbf{V}_{\bigcdot}, \boldsymbol{\beta}_{\bigcdot} &\sim \hbox{Bernoulli}(\Phi(\mathbf{V}^*\boldsymbol{\beta}_{\bigcdot})) \\
    \mathbf{z}|\X_{\bigcdot}, \mathbf{V}, \mathbf{V}_{\bigcdot}, \boldsymbol{\beta}_{\bigcdot} &\sim \hbox{Normal} (\mathbf{V}^*\boldsymbol{\beta}_{\bigcdot}, \mathbf{I}_{n\times n})\\
    \X_{\bigcdot} &| \mathbf{U}_{\bigcdot}, \mathbf{V}, \mathbf{W}_{\bigcdot}, \mathbf{V}_{\bigcdot} \sim \prod_{s=1}^q \prod_{i=1}^n \prod_{j=1}^{p_s} \hbox{Normal}\left(\X_s[j,i]| \mathbf{U}_s[j,\cdot]\mathbf{V}[i,\cdot]^T + \mathbf{W}_s[j,\cdot] \mathbf{V}_s[i,\cdot]^T, \sigma^2_s \right) \\
    \mathbf{U}_s[j,\cdot] &\sim \hbox{Normal}(\boldsymbol0, \lambda^{-1} \mathbf{I}_{r\times r}) \\
    \mathbf{V}[i,\cdot] &\sim \hbox{Normal}(\boldsymbol0, \lambda^{-1} \mathbf{I}_{r\times r}) \\
    \mathbf{W}_s[j,\cdot] &\sim \hbox{Normal}(\boldsymbol0, \lambda_s^{-1} \mathbf{I}_{r_s\times r_s}) \\
    \mathbf{V}_s[i,\cdot] &\sim \hbox{Normal}(\boldsymbol0, \lambda_s^{-1} \mathbf{I}_{r_s\times r_s}) \\
    \boldsymbol\beta_{\bigcdot} &\sim \hbox{Normal}\left(\boldsymbol0, \boldsymbol{\Sigma}_{\beta} \right)
\end{split}
\end{align}
Prior hyperparameters, $\lambda^{-1}$, $\lambda^{-1}_s$, and $\boldsymbol{\Sigma}_{\beta}$ are chosen as described in the main manuscript. As before, we assume the data sources, $\mathbf{X}_s$, are scaled to have overall variance $1$, i.e. $\sigma^2_s = 1$ for $s=1,\dots,q$. Imputation can be done in the same way described in the main manuscript and all conditional posteriors are described below.

\subsection{Conditional Posteriors with Binary Outcome}

Conditional posterior for $\mathbf{V}$: if a binary response vector, $\mathbf{y}$, is given, we sample from the posterior distribution for $\mathbf{V}$ by sampling from 

\begin{equation}\label{eq:BSFP_con_pos_V_bin}
P(\mathbf{V}[i,]\mid\mathbf{X}_{\bigcdot}, \mathbf{U}_{\bigcdot}, \mathbf{W}_{\bigcdot}, \mathbf{V}_{\bigcdot}, \{\sigma^2_s\}_{s=1}^q, \lambda^{-1}, \{\lambda^{-1}_{s} \}_{s=1}^q, \mathbf{y}, \mathbf{z}, \boldsymbol{\beta}_{\bigcdot}) = \hbox{Normal} \left(\mathbf{B}_V \mathbf{b}_V, \mathbf{B}_V \right)
\end{equation}
where:
\begin{align*}
    \mathbf{B}_V^{-1} &= \begin{pmatrix} \mathbf{U_{\bigcdot}} \\ \boldsymbol{\beta}_{\,joint}^T \end{pmatrix}^T \begin{pmatrix} \mathbf{\Sigma} & \boldsymbol{0} \\ \boldsymbol{0} & 1 \end{pmatrix}^{-1}\begin{pmatrix} \mathbf{U_{\bigcdot}} \\ \boldsymbol{\beta}_{\,joint}^T \end{pmatrix} + \frac{1}{\lambda^{-1}} \mathbf{I}_{r\times r} \\
    \mathbf{b}_V &= \begin{pmatrix} \mathbf{U_{\bigcdot}} \\ \boldsymbol{\beta}_{\,joint}^T \end{pmatrix}^T \begin{pmatrix} \mathbf{\Sigma} &  \boldsymbol{0}\\ \boldsymbol{0} & 1 \end{pmatrix}^{-1} \begin{pmatrix}
    (\mathbf{X}_{\bigcdot} - \mathbf{W}_{\bigcdot} \mathbf{V}_{\bigcdot}^T)[,i] \\
    (\mathbf{z} -\beta_0 - \sum_{s=1}^q \mathbf{V}_s \boldsymbol{\beta}_{indiv,s})[i,]
    \end{pmatrix}
\end{align*}

\noindent 
Conditional Posterior for $\mathbf{U}_{\bigcdot}$: for $s=1,\dots, q$ and $j = 1,\dots, p_s$, we can sample from the posterior distribution for $\mathbf{U}_s$ by sampling from 

\begin{equation}\label{eq:con_pos_U_bin}
P(\mathbf{U}_s[j,]\mid\mathbf{X}_{\bigcdot}, \mathbf{V}, \mathbf{W}_{\bigcdot}, \mathbf{V}_{\bigcdot}, \{\sigma^2_s\}_{s=1}^q, \lambda^{-1}, \{\lambda^{-1}_{s} \}_{s=1}^q, \mathbf{y}, \mathbf{z}, \boldsymbol{\beta}_{\bigcdot}) = \hbox{Normal}(\mathbf{B}_{U_s} \mathbf{b}_{U_s}, \mathbf{B}_{U_s})     
\end{equation}
where:
\begin{align*}
    \mathbf{B}_{U_s}^{-1} &= \frac{1}{\sigma^2_s} \mathbf{V^T} \mathbf{V} + \frac{1}{\lambda^{-1}} \mathbf{I}_{r\times r} \\
    \mathbf{b}_{U_s} &= \frac{1}{\sigma^2_s} \mathbf{V}^T (\mathbf{X}_s - \mathbf{W}_s \mathbf{V}_s^T)[j,]
\end{align*}
where $\mathbf{\Sigma}_s = \sigma^2_s \mathbf{I}_{p_s \times p_s}$. \\

\noindent 
Conditional Posterior for $\mathbf{V}_s$: If a binary response vector, $\mathbf{y}$, is given, for $s=1,\dots,q$ and $i=1,\dots,n$ we can sample from the posterior distribution of $\mathbf{V}_s$ by sampling from:

\begin{equation}\label{eq:BSFP_con_pos_Vs_bin}
P(\mathbf{V}_s[i,]\mid\mathbf{X}_{\bigcdot}, \mathbf{V}, \mathbf{W}_{\bigcdot}, \mathbf{U}_{\bigcdot}, \{\sigma^2_s\}_{s=1}^q, \lambda^{-1}, \{\lambda^{-1}_{s} \}_{s=1}^q, \mathbf{y}, \mathbf{z}, \boldsymbol{\beta}_{\bigcdot}) = \hbox{Normal}(\mathbf{B}_{V_s} \mathbf{b}_{V_s}, \mathbf{B}_{V_s}) \\    
\end{equation}
where:
\begin{align*}
    \mathbf{B}_{V_s}^{-1} = \begin{pmatrix}
    \mathbf{W}_s \\
    \boldsymbol{\beta}_{indiv,s}
    \end{pmatrix}^T \begin{pmatrix}
    \mathbf{\Sigma}_s & \boldsymbol{0} \\
    \boldsymbol{0} & 1
    \end{pmatrix}^{-1}\begin{pmatrix}
    \mathbf{W}_s \\
    \boldsymbol{\beta}_{indiv,s}
    \end{pmatrix} + \frac{1}{\lambda^{-1}_{s}} \mathbf{I}_{r_s\times r_s} \\
    \mathbf{b}_{V_s} = \begin{pmatrix}
    \mathbf{W}_s \\
    \boldsymbol{\beta}_{indiv,s}
    \end{pmatrix}^T \begin{pmatrix}
    \mathbf{\Sigma}_s & \boldsymbol{0} \\
    \boldsymbol{0} & 1
    \end{pmatrix}^{-1} \begin{pmatrix}
    (\mathbf{X}_s - \mathbf{U}_s \mathbf{V}^T)[,i] \\
    (\mathbf{z} - \beta_0 - \mathbf{V} \boldsymbol{\beta}_{joint} - \sum_{s' \neq s} \mathbf{V}_{s'} \boldsymbol{\beta}_{indiv,s'})[i,]
    \end{pmatrix}
\end{align*} 

\noindent 
Conditional Posterior for $\mathbf{W}_{\bigcdot}$: for $s=1,\dots, q$ and $j=1,\dots, p_s$, we can sample from the posterior distribution for $\mathbf{W}_s$ by sampling from 

\begin{equation}\label{eq:con_pos_Ws_bin}
P(\mathbf{W}_s[j,]\mid\mathbf{X}_{\bigcdot}, \mathbf{V}, \mathbf{V}_{\bigcdot}, \mathbf{U}_{\bigcdot}, \{\sigma^2_s\}_{s=1}^q, \lambda^{-1}, \{\lambda^{-1}_{s} \}_{s=1}^q,\mathbf{y}, \mathbf{z}, \boldsymbol{\beta}_{\bigcdot}) = \hbox{Normal}(\mathbf{B}_{W_s} \mathbf{b}_{W_s}, \mathbf{B}_{W_s})
\end{equation}
where:
\begin{align}
    \mathbf{B}_{W_s}^{-1} &= \frac{1}{\sigma^2_s}\mathbf{V}_s^T \mathbf{V}_s + \frac{1}{\lambda^{-1}_{s}} \mathbf{I}_{r_s\times r_s} \\
    \mathbf{b}_{W_s} &= \frac{1}{\sigma^2_s}\mathbf{V}_s^T (\mathbf{X}_s - \mathbf{U}_s \mathbf{V}^T)[j,]
\end{align}

\noindent 
Conditional Posterior for $\boldsymbol{\beta}_{\bigcdot}$: if $\mathbf{y}$ is binary, we can sample from the posterior for $\boldsymbol{\beta}_{\bigcdot}$ by sampling from:

\begin{equation}\label{eq:BSFP_con_pos_beta_bin}
P(\boldsymbol{\beta}_{\bigcdot}\mid\mathbf{X}_{\bigcdot}, \mathbf{V}, \mathbf{V}_{\bigcdot}, \mathbf{U}_{\bigcdot}, \{\sigma^2_s\}_{s=1}^q, \lambda^{-1}, \{\lambda^{-1}_{s} \}_{s=1}^q, \mathbf{y},\mathbf{z}) = \hbox{Normal}(\mathbf{B}_\beta \mathbf{b}_\beta, \mathbf{B}_\beta) 
\end{equation}
where:
\begin{align}
    \mathbf{B}_\beta^{-1} &=  \mathbf{V}^{*T} \mathbf{V}^{*} + \mathbf{\Sigma}_\beta^{-1} \\
    \mathbf{b}_\beta &= \mathbf{V}^{*T} \mathbf{z}
\end{align}

\noindent 
The conditional posterior of $\mathbf{z}$, the latent continuous response variable, is:

\begin{align*}
    z_i \mid\mathbf{X}_{\bigcdot}, \mathbf{V}, \mathbf{V}_{\bigcdot}, \mathbf{U}_{\bigcdot}, \{\sigma^2_s\}_{s=1}^q, \lambda^{-1}, \{\lambda^{-1}_{s} \}_{s=1}^q, \mathbf{y}, \boldsymbol{\beta}_{\bigcdot} &\sim \hbox{Normal}(\mathbf{V}^*[i,]\boldsymbol{\beta}_{\bigcdot}, 1) \tag{truncated at left by 0 if $y_i=1$}  \\
    z_i \mid\mathbf{X}_{\bigcdot}, \mathbf{V}, \mathbf{V}_{\bigcdot}, \mathbf{U}_{\bigcdot}, \{\sigma^2_s\}_{s=1}^q, \lambda^{-1}, \{\lambda^{-1}_{s} \}_{s=1}^q, \mathbf{y}, \boldsymbol{\beta}_{\bigcdot} &\sim \hbox{Normal}(\mathbf{V}^*[i,]\boldsymbol{\beta}_{\bigcdot}, 1) \tag{truncated at right by 0 if $y_i=0$}
\end{align*}

\section{Validation Simulation}\label{sup:validation}

We used simulations to validate that our in-house Gibbs sampling algorithm to estimate BSF and BSFP was properly sampling from the posterior distributions of the model parameters. In our simulations, we generated the entries in the model parameters, $\mathbf{V}$, $\mathbf{U}_{\bigcdot}$, $\mathbf{V}_{\bigcdot}$, $\mathbf{W}_{\bigcdot}$, $\boldsymbol{\beta}_{\bigcdot}$, and $\tau^2$, from their respective prior distributions. We generated $q=2$ sources of data, rank $r=1$ joint structure, and rank $r_s=1$ individual structures for $s=1,2$. Source $1$ was assumed to have $p_1=100$ features while source $2$ was assumed to have $p_2=150$ features. We generated $n=50$ samples, matched on both sources. We assumed the error variances of the two sources were $1$. We generated the entries in the joint factorization matrices, $\mathbf{V}$ and $\mathbf{U}_{\bigcdot}$, from a $\hbox{Normal}(0,1)$ distribution and the entries in the individual factorization matrices, $\mathbf{V}_{\bigcdot}$ and $\mathbf{W}_{\bigcdot}$ from a $\hbox{Normal}(0,1)$ distribution. We generated $\boldsymbol{\beta}_{\bigcdot}$ from a $\hbox{Normal}(0,\boldsymbol\Sigma_\beta)$ assuming a prior variance-covariance matrix of $\boldsymbol\Sigma_\beta = diag\{10, 1, \dots, 1 \}$, where $10$ reflects the prior variance on the intercept, and $1$ reflects the prior variance on the effects of each factor in explaining $\mathbf{y}$. We generated $\tau^2$ from an $\hbox{Inverse-Gamma}(1,1)$ distribution. We fixed the prior hyperparameters in the model fitting algorithm to match the hyperparameters used to generate $\mathbf{V}$, $\mathbf{U}_{\bigcdot}$, $\mathbf{V}_{\bigcdot}$, $\mathbf{W}_{\bigcdot}$, $\boldsymbol{\beta}_{\bigcdot}$, and $\tau^2$ so as to validate model performance. We considered including a continuous and binary outcome, with and without missingness, and inducing entrywise and blockwise missing in the data. We considered 30\%, 50\%, and 70\% levels of entrywise missingness and missingness in a continuous or binary response. We considered 10\% and 30\% levels of blockwise missingness. For computational reasons, we did not consider higher levels of blockwise missingness. We found that we had to vary the number of posterior sampling iterations depending on the condition to achieve proper coverage: without missingness, we considered 2000 posterior sampling iterations whereas with missingness, we considered 10000 or 20000 posterior sampling iterations. 

We monitored posterior coverage of the true underlying parameters using 95\% credible intervals, as well as relative squared error (RSE) and the width of the credible intervals to gauge model performance and uncertainty. We expect coverage to be around 95\%, the RSE to be close to $0$, and the width of the credible intervals to increase as we add additional sources of uncertainty, like missing values. 

The steps to our simulation were as follows. For $i=1,\dots, 100$:
\begin{enumerate}
    \item Generate true values for the underlying structures, $\mathbf{U}^{0}_{\bigcdot}$, $\mathbf{V}^{0}$, $\mathbf{V}^{0}_{\bigcdot}$, $\mathbf{W}^{0}_{\bigcdot}$ iid from a $\hbox{Normal}(0,1)$
    \item Generate each dataset using the underlying structure with additional random error simulated iid from a $\hbox{Normal}(0,1)$:

    \begin{align}
        \mathbf{X}_1 &= \mathbf{U}^{0}_1 \mathbf{V}^{0T} + \mathbf{W}^{0}_1 \mathbf{V}^{0T}_1 + \mathbf{E}_1 \\
        \mathbf{X}_2 &= \mathbf{U}^{0}_2 \mathbf{V}^{0T} + \mathbf{W}^{0}_2 \mathbf{V}^{0T}_2 + \mathbf{E}_2 
    \end{align}
    
    \item Generated the true coefficients $\boldsymbol{\beta}^0 \sim \hbox{Normal}(\boldsymbol0, \boldsymbol\Sigma_\beta)$ where $\boldsymbol\Sigma_\beta = diag\{10, 1, \dots, 1 \}$. 
    \begin{itemize}
        \item For a continuous response, simulate $\tau^{20} \sim \hbox{Inverse-Gamma}(1,1)$. 
        \begin{itemize}
            \item Then generate $\mathbf{y} \sim \hbox{Normal}(\mathbf{V}^{*0} \boldsymbol\beta^0, \tau^{20} I_{n\times n})$ where $\mathbf{V}^{*0} = \begin{pmatrix} \boldsymbol1 & \mathbf{V}^0 & \mathbf{V}^0_1 & \mathbf{V}^0_2 \end{pmatrix}$.
        \end{itemize}
        
        \item For a binary response, simulate $\mathbf{y} \sim \hbox{Bernoulli}_n(\Phi(\mathbf{V}^{*0} \boldsymbol\beta^0))$ where $\Phi(\cdot)$ refers to the cumulative distribution function of the standard normal distribution
    \end{itemize}
    
    \item Under conditions that required missingness in the data, we varied the percentage of missing entries to be 30\%, 50\%, and 70\% for entrywise missing and 10\% and 30\% for blockwise missing.
    
    \item Under conditions that required missingness in the response, we considered 30\%, 50\%, and 70\% of responses (both continuous and binary) to be missing. 
    
    \item Run the model fitting algorithm for $2000$ iterations with a $1000$ iteration burn-in. For higher amounts of missingness, we increased the number of iterations to $10000$ or $20000$ and used a $5000$ or $10000$ iteration burn-in, respectively.
    
    \item Calculate the 95\% credible intervals for the estimated underlying joint and individual structures using the samples from each Gibbs sampling iteration, $t$, $\mathbf{J}^{(t)}_s = \mathbf{U}^{(t)}_s\mathbf{V}^{(t)T}$ and  $\mathbf{A}^{(t)}_s = \mathbf{W}^{(t)}_s\mathbf{V}^{(t)T}_s$ for $s=1,2$ and $\mathbb{E}(\mathbf{y}|\mathbf{X}_{\bigcdot})^{(t)} = \mathbf{V}^{*(t)}\boldsymbol\beta^{(t)}_{\bigcdot}$ for continuous data and $\mathbb{E}(\mathbf{y}|\mathbf{X}_{\bigcdot})^{(t)} = \Phi(\mathbf{V}^{*(t)}\boldsymbol\beta^{(t)}_{\bigcdot})$ for binary data. Calculate the coverage, credible interval (CI) width, and RSE. Evaluate these metrics separately for observed and missing values. We averaged coverage for joint and individual structures between the two datasets.
\end{enumerate}

The results averaged across simulation replications can be found in Table~\ref{tab:validation_results}. We saw coverage rates around 95\%. RSEs were generally close to 0 but increased with increasing levels of missingness. Credible interval (CI) widths increased with additional uncertainty due to imputing missing values. Coverage levels slightly below $0.95$ are likely due to an insufficient number of Gibbs sampling iterations and would be remedied by increasing this. 

\begin{table}[H]
\centering
\caption{Simulation results to validate model performance. $\mathbf{J}_{\bigcdot}$ and $\mathbf{A}_{\bigcdot}$ refer to results for the estimated joint and individual structures, respectively. If missingness was considered, $\mathbf{J}^{(m)}_{\bigcdot}$ and $\mathbf{A}^{(m)}_{\bigcdot}$ refer to joint and individual structures corresponding to missing entries. $\mathbb{E}(\mathbf{y}|\mathbf{X}_{\bigcdot})$ refers to results for estimating the conditional expectation of the response vector, if a response was considered. $\mathbb{E}\left(\mathbf{y}^{(m)}|\mathbf{X}_{\bigcdot}\right)$ refers to results for the conditional expectation of the response vector corresponding to missing entries. $\tau^2$ refers to results for the estimated error variance in $\mathbf{y}$. Cells are left blank in the table if the corresponding parameter was not estimated. }
\label{tab:validation_results}
\resizebox{\columnwidth}{!}{
\begin{tabular}{ccccccccc}
  \hline
Condition & Metric & $\mathbf{J}_{\bigcdot}$ & $\mathbf{A}_{\bigcdot}$ & $\mathbf{J}^{(m)}_{\bigcdot}$ & $\mathbf{A}^{(m)}_{\bigcdot}$ & $\mathbb{E}(\mathbf{y}|\mathbf{X}_{\bigcdot})$ & $\mathbb{E}\left(\mathbf{y}^{(m)}|\mathbf{X}_{\bigcdot}\right)$ & $\tau^2$\\
  \hline
 & Coverage & 0.9496 & 0.9507 &  &  &  &  &  \\ 
  No Response, No Missing & RSE & 0.0308 & 0.0345 &  &  &  &  &  \\ 
& CI Width & 0.5717 & 0.6347 &  &  &  &  &  \\ 
\hline
& Coverage & 0.9455 & 0.9459 &  &  & 0.9448 &  & 0.9700 \\ 
  Continuous Response & RSE & 0.0313 & 0.0368 &  &  & 0.0410 &  & 0.0405 \\ 
& CI Width & 0.5744 & 0.6343 &  &  & 1.7562 &  & 4.1543 \\ 
\hline
& Coverage & 0.9434 & 0.9446 &  &  & 0.9522 &  &  \\ 
  Binary Response & RSE & 0.0377 & 0.0423 &  &  & 0.0691 &  &  \\ 
& CI Width & 0.9440 & 1.0270 &  &  & 0.3921 &  &  \\ 
\hline
& Coverage & 0.9473 & 0.9463 & 0.9467 & 0.9464 &  &  &  \\ 
  30\% Entrywise Missing & RSE & 0.0419 & 0.0501 & 0.0455 & 0.0547 &  &  &  \\ 
& CI Width & 0.6697 & 0.7444 & 0.6867 & 0.7671 &  &  &  \\ 
\hline
& Coverage & 0.9461 & 0.9459 & 0.9456 & 0.9459 &  &  &  \\ 
  50\% Entrywise Missing & RSE & 0.0572 & 0.0682 & 0.0639 & 0.0772 &  &  &  \\ 
& CI Width & 0.7890 & 0.8740 & 0.8190 & 0.9122 &  &  &  \\ 
\hline
& Coverage & 0.9464 & 0.9434 & 0.9456 & 0.9435 &  &  &  \\ 
  70\% Entrywise Missing & RSE & 0.0971 & 0.1114 & 0.1177 & 0.1352 &  &  &  \\ 
& CI Width & 1.0181 & 1.1275 & 1.0892 & 1.2142 &  &  &  \\ 
\hline
& Coverage & 0.9463 & 0.9457 &  &  & 0.9404 &  & 0.9400 \\ 
  30\% Continuous Response Missing & RSE & 0.0311 & 0.0365 &  &  & 0.0764 &  & 0.0891 \\ 
& CI Width & 0.5717 & 0.6323 &  &  & 2.1015 &  & 4.7370 \\ 
\hline
& Coverage & 0.9475 & 0.9466 &  &  & 0.9561 & 0.9523 & 0.9600 \\ 
  50\% Continuous Response Missing & RSE & 0.0352 & 0.0389 &  &  & 0.1944 & 0.2207 & 0.1184 \\ 
& CI Width & 0.5756 & 0.6374 &  &  & 2.1528 & 2.3872 & 3.9160 \\ 
\hline
& Coverage & 0.9483 & 0.9490 &  &  & 0.9456 & 0.9456 & 0.9500 \\ 
  70\% Continuous Response Missing & RSE & 0.0319 & 0.0358 &  &  & 0.2519 & 0.2778 & 0.1722 \\ 
& CI Width & 0.5791 & 0.6380 &  &  & 2.6876 & 3.0484 & 6.4290 \\ 
\hline
& Coverage & 0.9428 & 0.9417 &  &  & 0.9370 & 0.9150 &  \\ 
  30\% Binary Response Missing & RSE & 0.0389 & 0.0431 &  &  & 0.0722 & 0.0950 &  \\ 
& CI Width & 0.9469 & 1.0278 &  &  & 0.4525 & 0.5071 &  \\ 
\hline
& Coverage & 0.9488 & 0.9483 &  &  & 0.9387 & 0.9428 &  \\ 
  50\% Binary Response Missing & MSE & 0.0317 & 0.0362 &  &  & 0.0973 & 0.1207 &  \\ 
& CI Width & 0.5754 & 0.6399 &  &  & 0.4229 & 0.4726 &  \\ 
\hline
& Coverage & 0.9475 & 0.9472 &  &  & 0.9275 & 0.9231 &  \\ 
  70\% Binary Response Missing & MSE & 0.0323 & 0.0376 &  &  & 0.2482 & 0.2227 &  \\ 
& CI Width & 0.5773 & 0.6385 &  &  & 0.5004 & 0.5844 &  \\
\hline
& Coverage & 0.9473 & 0.9459 & 0.9493 & 0.9313 &  &  &  \\ 
  10\% Blockwise Missingness & RSE & 0.0345 & 0.0413 & 0.0500 & 1.0738 &  &  &  \\ 
& CI Width & 0.6101 & 0.6662 & 0.6452 & 3.1696 &  &  &  \\ 
\hline
& Coverage & 0.9518 & 0.9507 & 0.9519 & 0.9351 &  &  &  \\ 
  30\% Blockwise Missingness & RSE & 0.0451 & 0.0463 & 0.0498 & 1.0403 &  &  &  \\ 
& CI Width & 0.6885 & 0.7299 & 0.7406 & 3.2038 &  &  &  \\ 
   \hline
\end{tabular}}
\end{table}

\section{Initializing with a Continuous Response}\label{sup:init_with_y}

Unlike BSF, BSFP is not initialized at the true posterior mode of the model as the initialization does not account for the outcome. We now describe a formulation of BSFP where we included a continuous response, $\mathbf{y}$, in the initialization. This amounts to treating $\mathbf{y}$ as an additional source of data. The challenge with including a response vector in the initialization is that the UNIFAC and BSFP frameworks assume the error variances across data sources are equal, typically $1$. For matrix data sources, this is easily accommodated. We may estimate the error variance using the MAD estimator and scale the sources accordingly. However, the MAD estimator is premised on the random matrix having low-rank structure, which does not apply to a vector. Thus, we cannot apply the MAD estimator to $\mathbf{y}$. The challenge with estimating the error variance in $\mathbf{y}$ is that the data are high-dimensional, so standard linear models are not appropriate.

Instead, we considered estimating the error variance using a lasso model. We included all observed features across the sources as predictors. We then used the natural lasso estimator, proposed by \cite{yu2019estimating}, to estimate the error variance in $\mathbf{y}$. We scaled $\mathbf{y}$ by the resulting estimated error standard deviation to obtain a response vector with an error variance approximately equal to $1$. We then considered initializing the BSFP model with UNIFAC where $\mathbf{y}$ is treated as an additional source of data. In our simulations, we generate $\mathbf{y}$ with a true error variance of $1$.

We compare the original formulation of BSFP to two alternative versions of this model. In one version, we consider BSFP with $\mathbf{y}$ included as an additional source during initialization. Since $\mathbf{y}$ is generated with an assumed error variance of $1$, we do not do any additional scaling of $\mathbf{y}$ and assume the error variance in $\mathbf{y}$ equals $1$. We refer to this model as ``BSFP (Init with Y, Fix Y Var)." In the other version, termed ``BSFP (Init with Y, Estimate Y Var).", we consider BSFP with $\mathbf{y}$ included as an additional source during initialization but we scale $\mathbf{y}$ using the natural lasso estimator of the error standard deviation. After scaling, we assume the error variance in $\mathbf{y}$ is equal to $1$.

The hyperparameters of the priors used in BSFP (Init with Y, Fix Y Var) and BSFP (Init with Y, Estimate Y Var) were fixed at the same values. The choice of hyperparameters was motivated by the nuclear norm-penalized objective. Invoking the notation used in the main manuscript, we fixed $\lambda^{-1} = (\sqrt{n}+\sqrt{p}+1)^{-1}$ and $\lambda_s^{-1} = (\sqrt{n} + \sqrt{p_s}+ 1)^{-1}$, which nearly matches that used in the original formulation of BSFP with the addition of the $+1$ to account for $\mathbf{y}$ in the initialization. We also fixed $\boldsymbol\Sigma_\beta = diag\{10, (\sqrt{n}+\sqrt{p}+1)^{-1} \mathbf{I}_{r\times r}, (\sqrt{n}+\sqrt{p_1}+1)^{-1} \mathbf{I}_{r_1\times r_1}, \dots,  (\sqrt{n}+\sqrt{p_q}+1)^{-1} \mathbf{I}_{r_q\times r_q}\}$. This is in contrast to the prior variance-covariance matrix of $\boldsymbol\beta_{\bigcdot}$ used in BSFP, which was $\boldsymbol\Sigma_\beta = diag\{10, 1, 1, \dots, 1 \}$.

The simulation set-up matches that which was described in the model comparison simulation in the main manuscript. We generated two sources of data with joint rank fixed at $1$ and individual ranks each fixed at $1$. We assumed the error variance in the data sources was equal to $1$ and the error variance in the response was equal to $1$. We generated $100$ features in each source and $200$ samples matched across sources which were split into a training and test set. The data sources and response were generated from our assumed model with prior variances on the structure components fixed at $1$. Each model was fit on the full training $\mathbf{X}_{\bigcdot}$ with only access to $\mathbf{y}^{train}$. Prediction was assessed using the held-out $\mathbf{y}^{test}$.

\begin{figure}[H]
    \centering
    \includegraphics[scale=0.3]{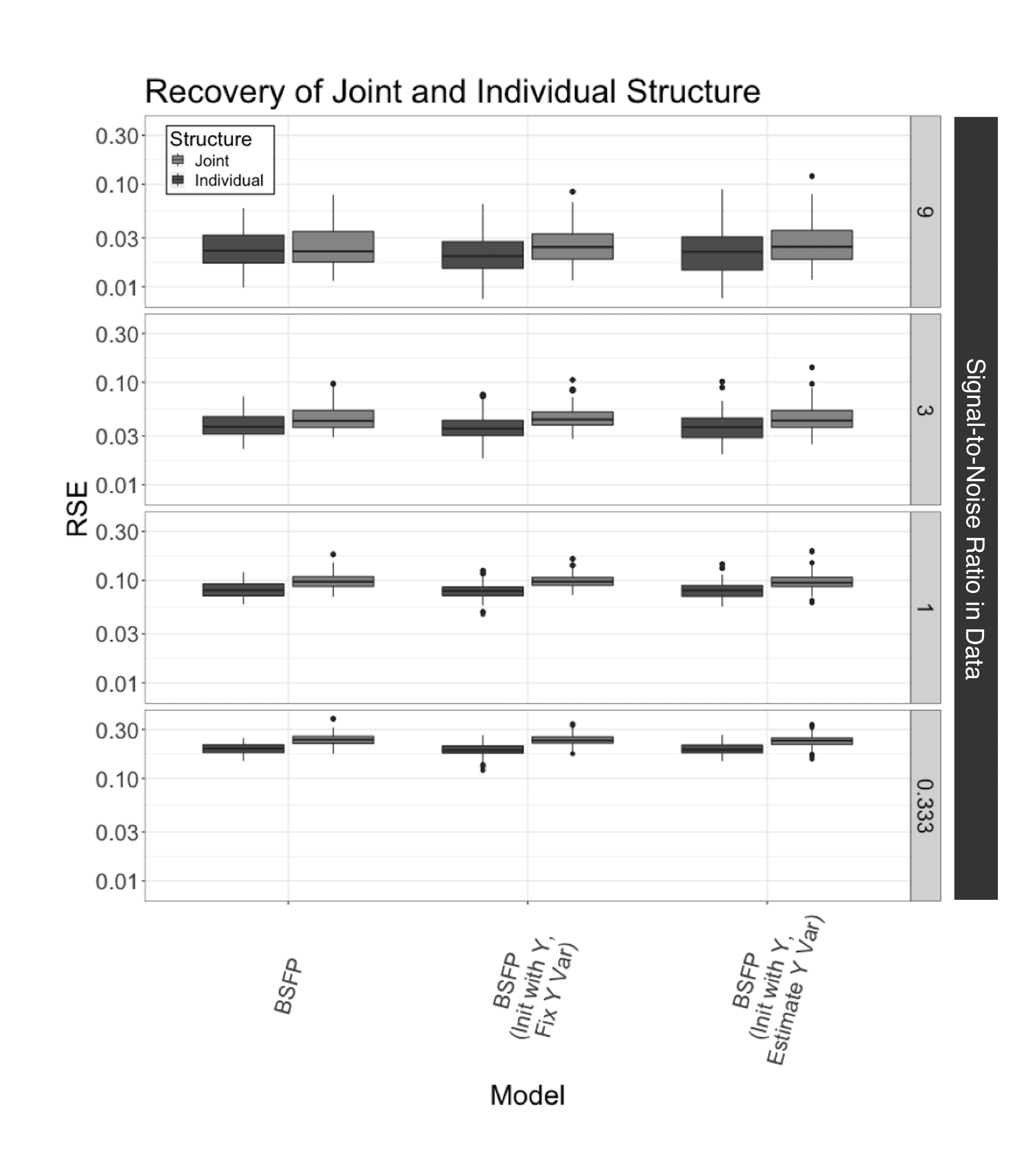}
    \caption{Comparing the recovery of the underlying structure between the original formulation of BSFP to two additional formulations which were initialized with $\mathbf{y}$ as an additional source. RSE levels close to 0 reflect better performance. }
    \label{fig:init_with_y_structure}
\end{figure}

\begin{figure}[H]
    \centering
    \includegraphics[scale=0.3]{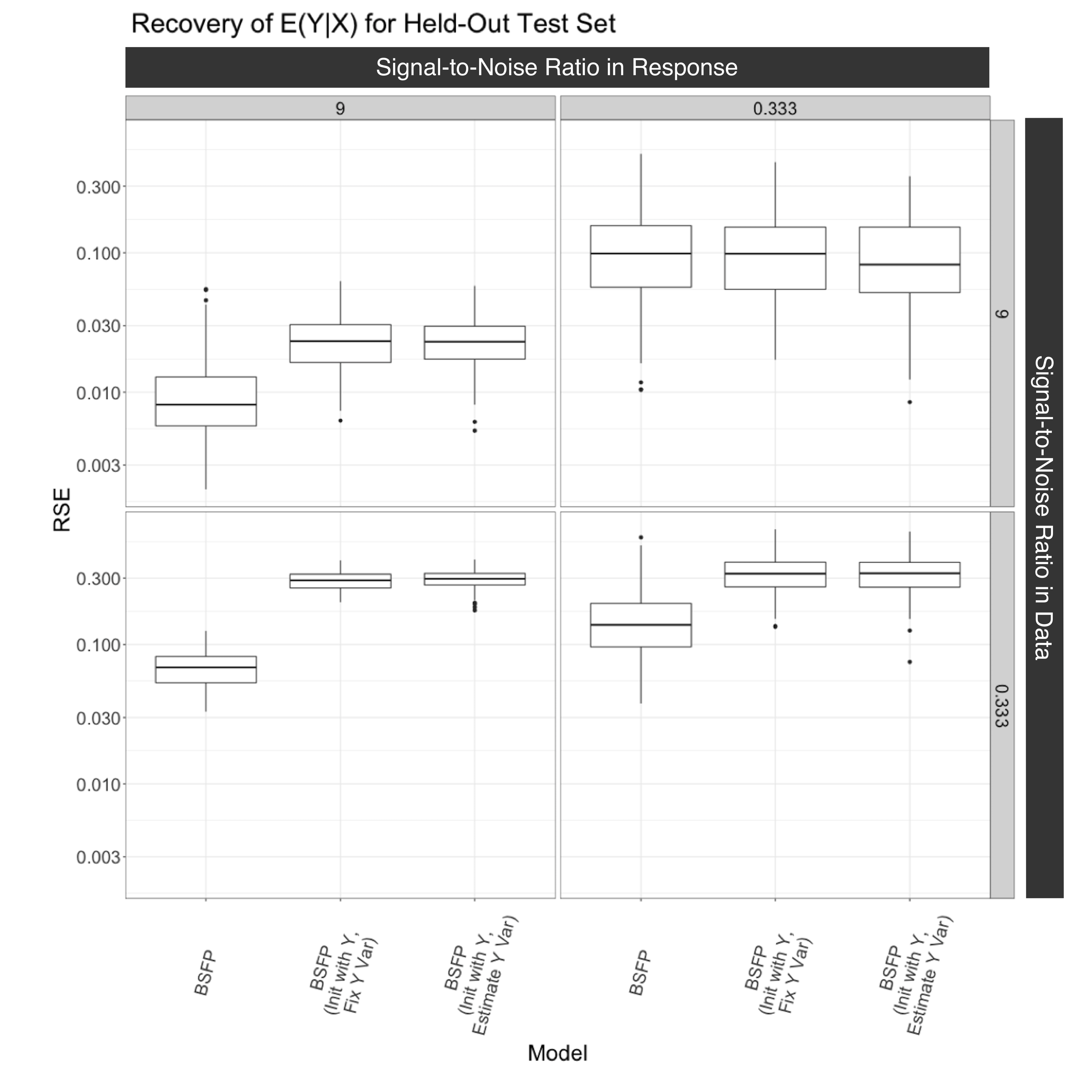}
    \caption{Comparing the recovery of $\mathbb{E}(\mathbf{y}^{test}|\mathbf{X}_{\bigcdot})$ between the original formulation of BSFP to two additional formulations which were initialized with $\mathbf{y}$ as an additional source. RSE values closer to $0$ reflect better performance. We select only the highest and lowest s2n ratios for space considerations. }
    \label{fig:init_with_y_prediction}
\end{figure}

\begin{figure}[H]
    \centering
    \includegraphics[scale=0.3]{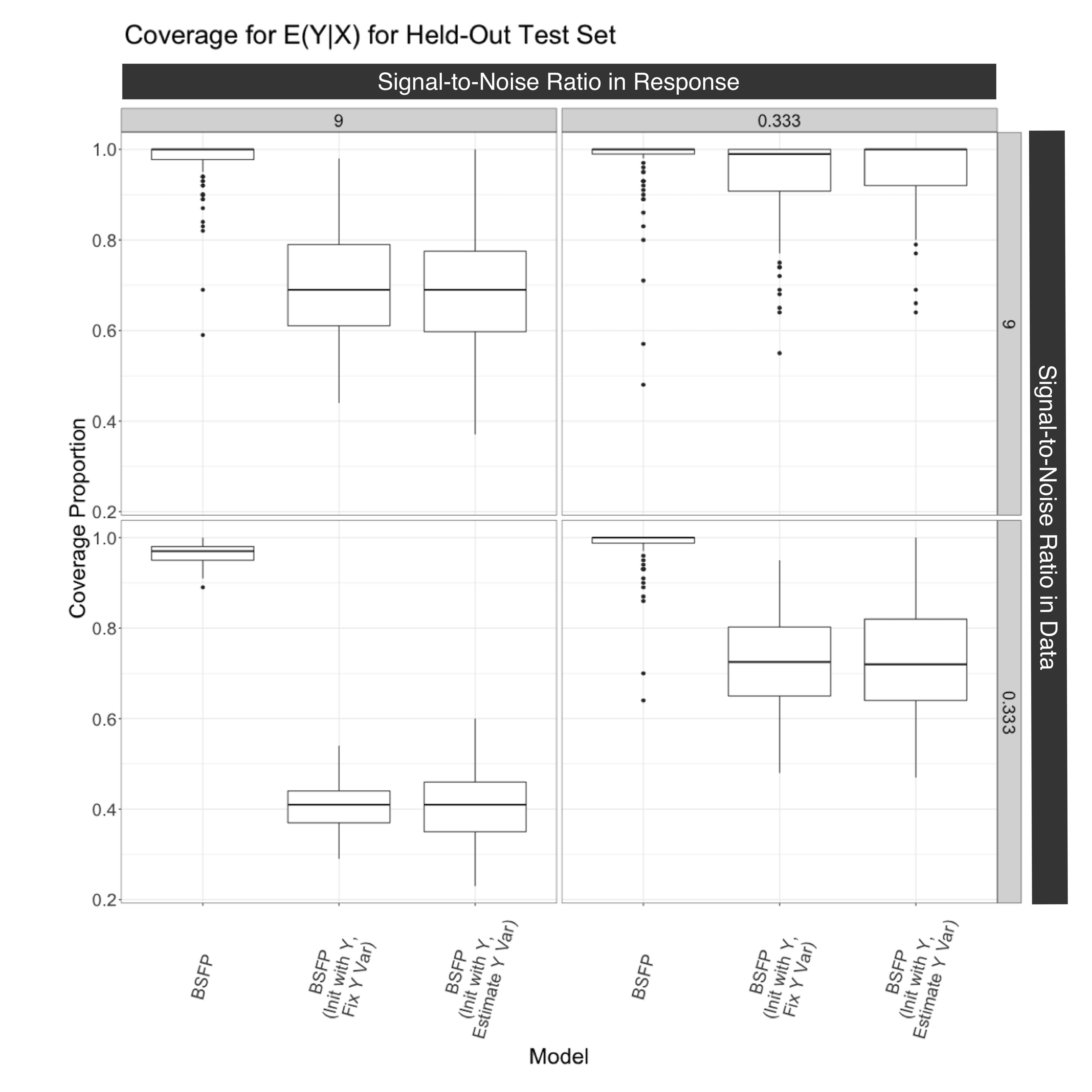}
    \caption{Comparing the coverage of $\mathbb{E}(\mathbf{y}^{test}|\mathbf{X}_{\bigcdot})$ comparing the original formulation of BSFP to two additional formulations which were initialized with $\mathbf{y}$ as an additional source. We select only the highest and lowest s2n ratios for space considerations.}
    \label{fig:init_with_y_coverage}
\end{figure}

The results for recovery of the underlying structure are shown in Figure~\ref{fig:init_with_y_structure}. All three versions of BSFP were comparable in recovering the underlying structure. There may have been a marginal benefit to initializing with $\mathbf{y}$ (slightly lower RSE, less variation in RSE across simulation replications), but the benefits are slim. The results for prediction of $\mathbf{y}$ are shown in Figure~\ref{fig:init_with_y_prediction}. We found comparable estimation of $\mathbb{E}(\mathbf{y}|\mathbf{X}_{\bigcdot})$ across all signal-to-noise levels, the differences in RSE were only marginal. The original formulation of BSFP, where the error variance in the response is estimated during model fitting, overall performed better. We saw larger differences in performance when we considered coverage of $\mathbb{E}(\mathbf{y}|\mathbf{X}_{\bigcdot})$ in Figure~\ref{fig:init_with_y_coverage}. Here, the original formulation of BSFP yielded nominal coverage.

\begin{figure}[H]
    \centering
    \includegraphics[scale=0.3]{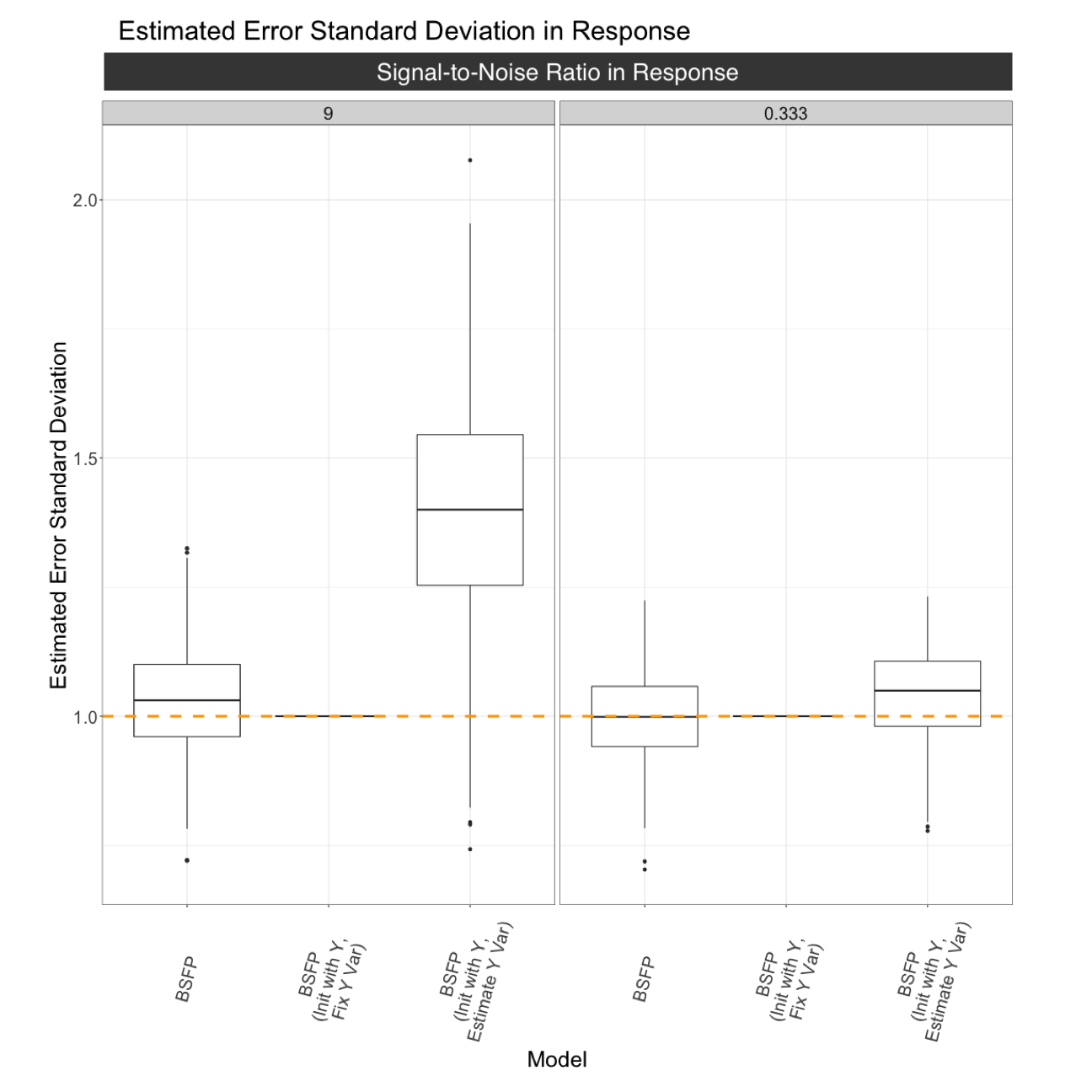}
    \caption{Estimated error standard deviation in $\mathbf{y}$ comparing the original formulation of BSFP to two additional formulations which were initialized with $\mathbf{y}$ as an additional source. Orange dashed line reflects the true error standard deviation in $\mathbf{y}$.}
    \label{fig:init_with_y_est_err_sd}
\end{figure}

The difference in predictive performance appears to be due to the specification of the predictive model for $\mathbf{y}$, as we know from Figure~\ref{fig:init_with_y_structure} all methods were able to accurately recover the underlying structure. Some performance differences may be due to the additional uncertainty in estimating the error standard deviation in $\mathbf{y}$. Under high signal in $\mathbf{y}$, we found that the natural lasso estimator overestimated the error standard deviation, as seen in Figure~\ref{fig:init_with_y_est_err_sd}. However, the biggest factor affecting model performance is likely the specification of the prior variance-covariance matrix for $\boldsymbol\beta_{\bigcdot}$. In BSFP (Init with Y, Fix Y Var) and BSFP (Init with Y, Estimate Y Var), these prior variances were fixed at low, informative values so as to match the posterior mode given by the UNIFAC objective. In BSFP, these prior variances were less informative and matched the data-generating model. The choice of hyperparameters for $\boldsymbol\beta_{\bigcdot}$ in BSFP (Init with Y, Fix Y Var) and BSFP (Init with Y, Estimate Y Var) was motivated by random matrix theory results for matrices which do not translate well when considering vectors, yielding an overly-restrictive prior. This highlights the challenge of incorporating $\mathbf{y}$ into the initialization using the nuclear norm-penalized objective. 

\section{Imputation Simulation}\label{sup:imputation}

\subsection{Results for Rank 3}

In the main manuscript, we describe a simulation to characterize the imputation performance of BSF under a relatively-low rank of $3$ and a relatively-higher rank of $15$. Here, we focus on results when the overall rank of the simulated data is $3$. All data-generating steps match those given in the main manuscript. In addition to fixing the overall rank at $3$, we consider $4$ signal-to-noise levels ($9$, $3$, $1$, and $1/3$) and run each method for $100$ replications. Imputation accuracy was evaluated using the RSE. Results are averaged across sources and replications. 

\begin{figure}
    \centering
    \includegraphics[scale=0.2]{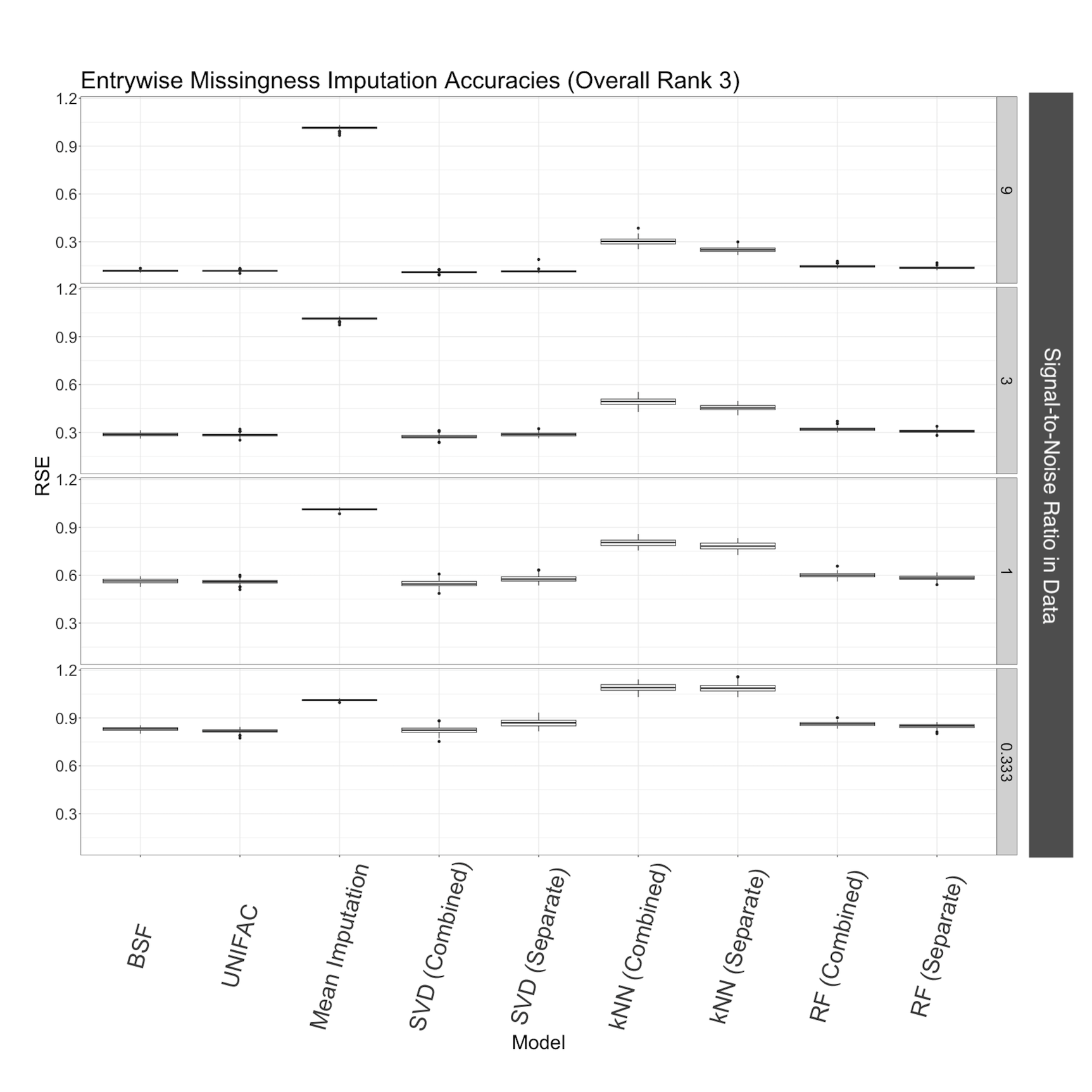}
    \caption{Imputation accuracy results under an assumed overall rank of $3$ with entrywise missing data. Models are compared based on the relative squared error (RSE) of the unobserved values compared to the imputed values. For BSF, the RSE is calculated using the posterior mean of the imputed values.}
    \label{fig:entrywise_imputation_rank3}
\end{figure}

In Figure~\ref{fig:entrywise_imputation_rank3}, we show results under entrywise missingness. Across all s2n levels, BSF, UNIFAC, SVD, and RF were competitive in imputing missing values. It is not surprising that the SVD performs well, as it was fit with an assumed rank of $4$ and likely recovers well the underlying structure without overfitting. All methods showed very little variance in their performance across the simulation replications. 

\begin{figure}
    \centering
    \includegraphics[scale=0.2]{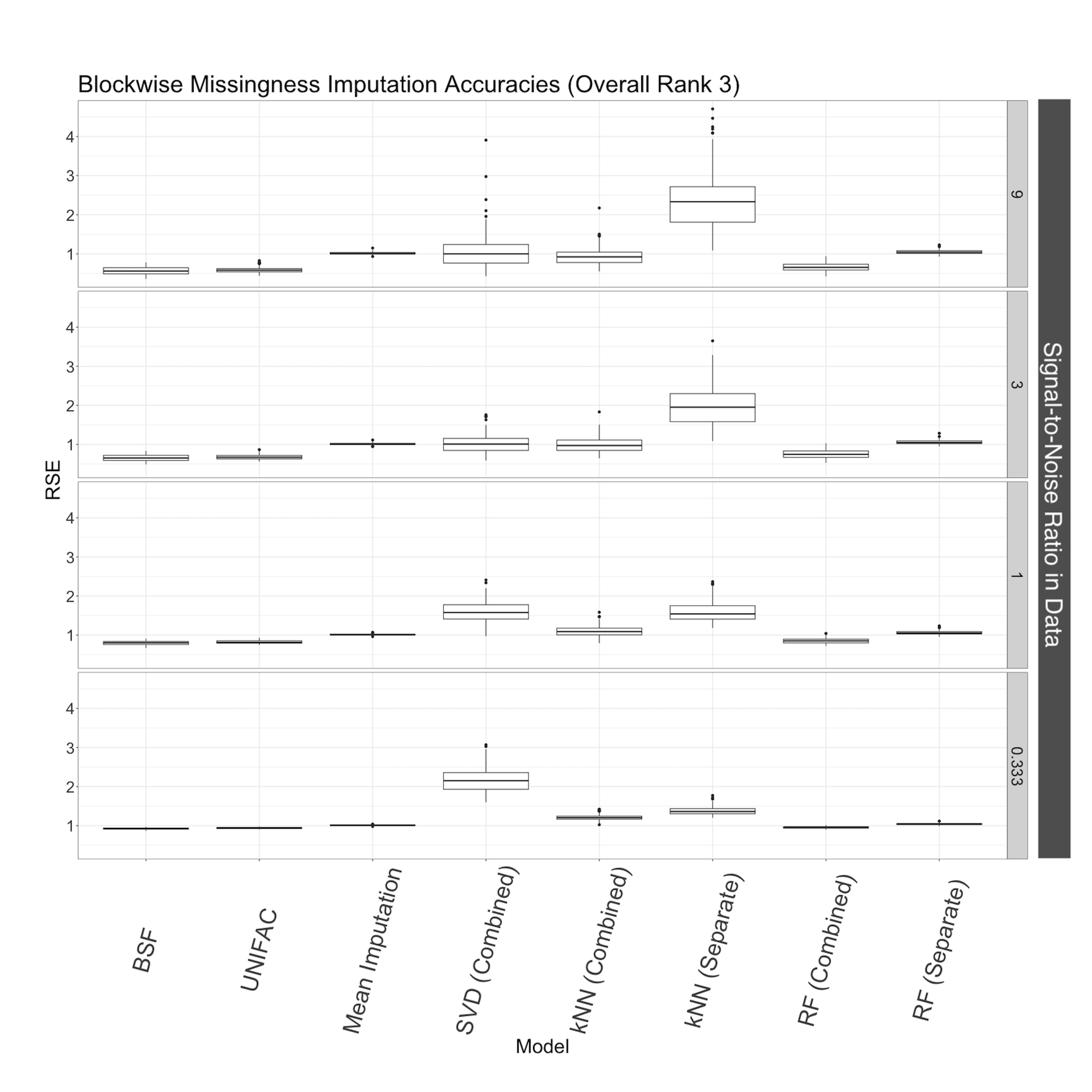}
    \caption{Imputation accuracy results under an assumed overall rank of $3$ with blockwise missing data. Models are compared based on the relative squared error (RSE) of the unobserved values compared to the imputed values. For BSF, the RSE is calculated using the posterior mean of the imputed values.}
    \label{fig:blockwise_imputation_rank3}
\end{figure}

Results for blockwise missingness are given in Figure~\ref{fig:blockwise_imputation_rank3}. As described in the main manuscript, imputation methods applied to each source separately have no information to use in imputing unobserved samples. This explains why low-rank factorization methods like BSF and UNIFAC perform best. RF remains a strong alternative method when applied to both sources combined. 

\begin{figure}
    \centering
    \includegraphics[scale=0.2]{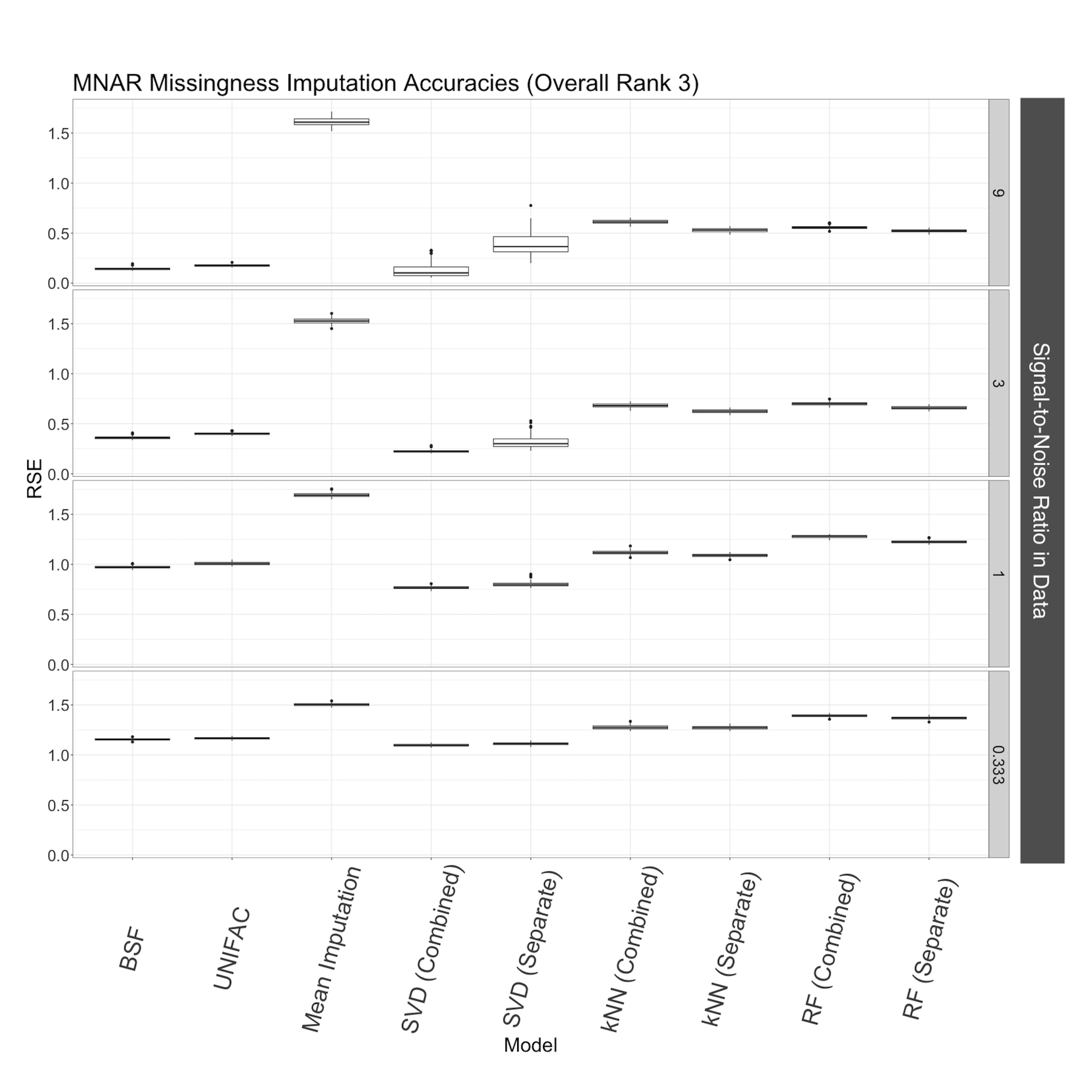}
    \caption{Imputation accuracy results under an assumed overall rank of $3$ with data missing-not-at-random (MNAR). Models are compared based on the relative squared error (RSE) of the unobserved values compared to the imputed values. For BSF, the RSE is calculated using the posterior mean of the imputed values.}
    \label{fig:MNAR_imputation_rank3}
\end{figure}

Results under MNAR are given in Figure~\ref{fig:MNAR_imputation_rank3}. As with an overall rank of $15$, BSF and UNIFAC perform very well under this missingness assumption under high signal. SVD applied to the combined sources is provides competitive imputation as an alternative method. 

\subsection{Imputation Uncertainty}

In this section, we discuss the uncertainty in imputing unobserved values using BSF in our imputation simulation. We focus here on the coverage and width of 95\% credible intervals based on imputing unobserved values using the posterior predictive distribution. We compare results across entrywise, blockwise, and missing-not-at-random conditions under different levels of signal in the data. 

In Tables~\ref{tab:imp_uncertainty_rank3} and ~\ref{tab:imp_uncertainty_rank15}, we study the uncertainty in imputing missing values under different levels of signal and different types of missingness using BSFP under an assumed rank of $3$ and an assumed rank of $15$. In both cases, we saw a similar pattern of uncertainty. In general, under entrywise missingness, the imputation accuracy (mean RSE) decreased as signal decreased, as it became more challenging to accurately recover the underlying structure. As signal decreased under entrywise missingness, we saw marginal declines in coverage and an overall decline in the average credible interval width. With lower signal, we tend to impute missing values using values close to $0$ and the overall drop in variance of these imputed values is reflected in the CI width. 

\begin{table}[H]
\centering
\caption{Uncertainty in imputed values when the data was generated with an assumed rank of 3.}
\label{tab:imp_uncertainty_rank3}
\begin{tabular}{ccccc}
  \hline
s2n & Missingness Type & Mean RSE & Mean Coverage & Mean CI Width \\ 
  \hline
9 & Entrywise & 0.1184 & 0.9673 & 4.6761 \\ 
  9 & Blockwise & 0.5603 & 0.6851 & 3.8032 \\ 
  9 & MNAR & 0.1423 & 0.7152 & 4.5820 \\ 
  \hline 
  3 & Entrywise & 0.2872 & 0.9469 & 4.1661 \\ 
  3 & Blockwise & 0.6537 & 0.7938 & 3.7263 \\ 
  3 & MNAR & 0.3603 & 0.5614 & 3.9091 \\ 
  \hline 
  1 & Entrywise & 0.5632 & 0.9340 & 3.9143 \\ 
  1 & Blockwise & 0.7977 & 0.8646 & 3.6758 \\ 
  1 & MNAR & 0.9713 & 0.3707 & 3.0652 \\ 
  \hline 
  0.333 & Entrywise & 0.8305 & 0.9272 & 3.7960 \\ 
  0.333 & Blockwise & 0.9269 & 0.9003 & 3.6409 \\ 
  0.333 & MNAR & 1.1564 & 0.3390 & 2.9427 \\ 
   \hline
\end{tabular}
\end{table}

\begin{table}[H]
\centering
\caption{Uncertainty in imputed values when the data was generated with an assumed rank of 15.}
\label{tab:imp_uncertainty_rank15}
\begin{tabular}{ccccc}
  \hline
s2n & Missingness Type & Mean RSE & Mean Coverage & Mean CI Width \\ 
  \hline
9 & Entrywise & 0.2261 & 0.9350 & 5.5594 \\ 
  9 & Blockwise & 0.6384 & 0.6283 & 4.2770 \\ 
  9 & MNAR & 0.5729 & 0.3897 & 6.3075 \\ 
  \hline 
  3 & Entrywise & 0.4602 & 0.9133 & 4.6483 \\ 
  3 & Blockwise & 0.7633 & 0.7656 & 4.0584 \\ 
  3 & MNAR & 0.8706 & 0.2846 & 4.5944 \\ 
  \hline 
  1 & Entrywise & 0.7903 & 0.9039 & 4.1939 \\ 
  1 & Blockwise & 0.9166 & 0.8533 & 3.9010 \\ 
  1 & MNAR & 1.3614 & 0.2599 & 3.3007 \\ 
  \hline 
  0.333 & Entrywise & 0.9773 & 0.9156 & 3.9481 \\ 
  0.333 & Blockwise & 0.9923 & 0.8997 & 3.7835 \\ 
  0.333 & MNAR & 1.2615 & 0.3140 & 3.0191 \\ 
   \hline
\end{tabular}
\end{table}

\bibliographystyle{plainnat}
\bibliography{bibliography}

\end{document}